\documentclass[12pt,a4paper]{article}
\usepackage{latexsym}
\usepackage{epsf}
\usepackage{graphicx}
\usepackage{color}
\def\be{\begin{equation}}
\def\ee{\end{equation}}
\def\ba{\begin{eqnarray}}
\def\ea{\end{eqnarray}}
\def\by{\left(\begin{array}}
\def\ey{\end{array}\right)}
\def\nnb{\nonumber \\}
\def\slash#1{\setbox0=\hbox{$#1$}#1\hskip-\wd0\dimen0=5pt\advance
       \dimen0 by-\ht0\advance\dimen0 by\dp0\lower0.5\dimen0\hbox
         to\wd0{\hss\sl/\/\hss}}

\renewcommand{\vec}[1]{\mbox{\boldmath$#1$}}

\begin{document}

\begin{center}
{\Large{Differential HBT Method for Binary Stars}}

\bigskip 

\small {Laszlo P. Csernai$^{1}$, Eirik S. Hatlen$^{1}$, 
and Sven Zschocke$^{1,2}$}
\end{center}

\footnotesize{
\begin{center}
$^{1}$ Department of Physics and Technology,\\
University of Bergen, Allegaten 55, 5007 Bergen, Norway \\
\vspace{0.2cm}
$^{2}$ Institute of Planetary Geodesy - Lohrmann-Observatory, \\
Dresden Technical University, Helmholtzstrasse 10, D-01069 Dresden, Germany\\
\end{center}
}
\normalsize



\begin{abstract}
Two-photon correlations in the thermal radiation field of a double-star system are studied.  
It is investigated how the differential Hanbury Brown and Twiss (HBT) approach 
can be used in order to determine orbital parameters of a binary star.  
\end{abstract}

\date{\today}

\newpage

\tableofcontents

\newpage

\section{Introduction}
\label{I}

Hanbury Brown and Twiss have discovered the remarkable fact that photons, emitted by some thermal light-source, tend 
to arrive at distant detectors in correlated pairs \cite{HBT2,HBT3}.  
As recognized by Hanbury Brown and Twiss, this HBT effect of photon correlation allows  
to determine the angular size of the thermal light-source.  
Soon afterwards, momentum-correlations among two pions created in heavy-ion collisions have   
been measured by Goldhaber et al. \cite{FirstHBTs}, which are a model-independent possibility in order to 
obtain information about the spatial size and the time-evolution of the hadronic fireball.  
Some typical two-pion correlation functions measured by experiments at CERN-SPS or at AGS-Brookhaven  
are shown in \cite{Experiment1} and \cite{Experiment2}, respectively.
Ever since, the HBT approach was used extensively for the study of relativistic heavy-ion reactions \cite{MLisa};  
for a historical review see \cite{Lisa}. Meanwhile, the method was  
extended not only to detect the size of the source of emission, but also
the speed of radial expansion \cite{Pratt84}, the rotational motion  
\cite{hydro1} and the turbulence \cite{hydro2} of the emitting source, also by means 
of the differential HBT method \cite{Differential_HBT_1,Differential_HBT_2,Differential_HBT_3}.  

\begin{figure}[!ht]
\begin{center}
\includegraphics[scale=0.45]{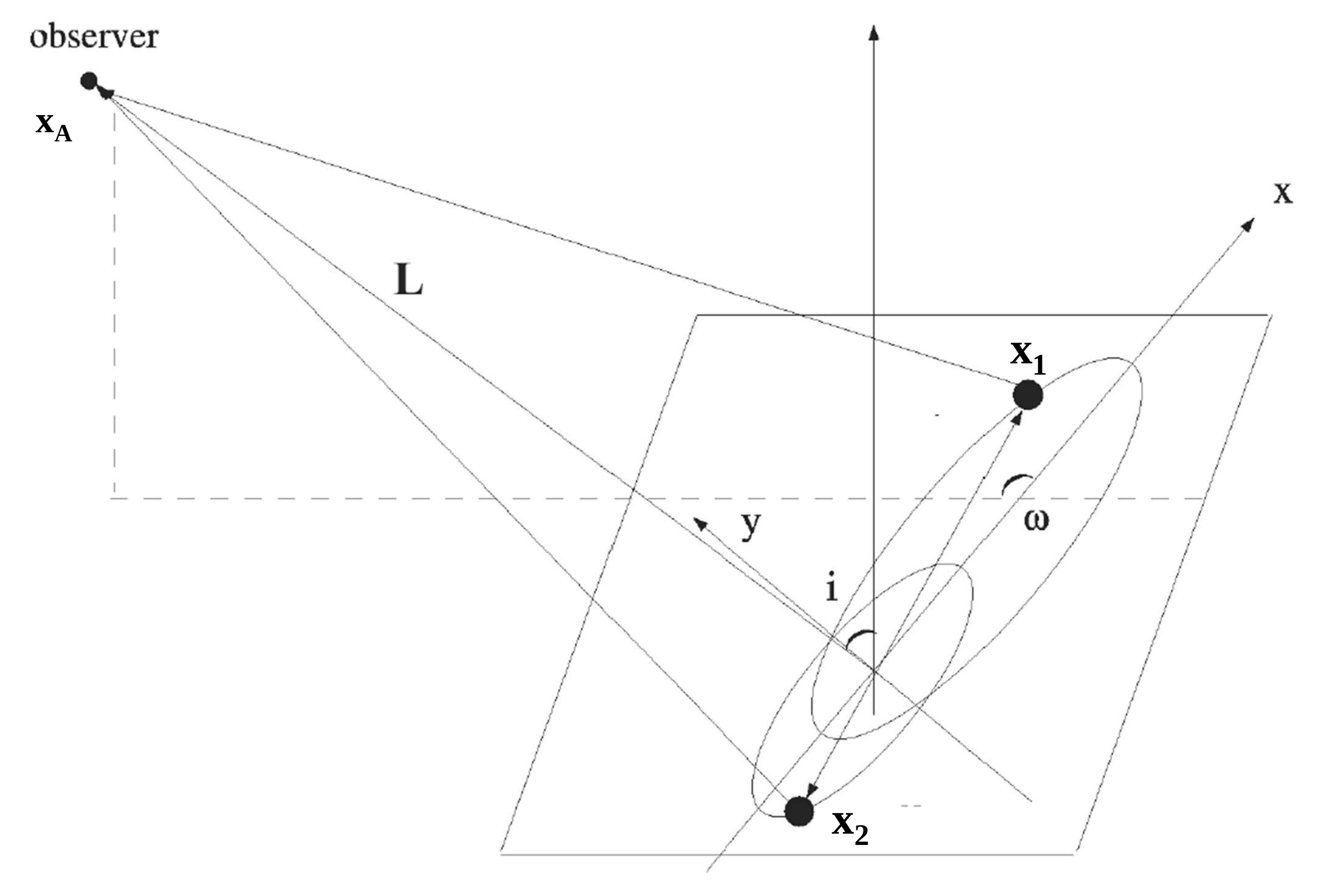}
\end{center}
\vskip -4mm
\caption{The orbit of a binary system is defined by seven orbital elements:
distance $\vec{L}$ between the center-of-mass of the binary star and the observer,  
semi-major axis $\vec x$, inclination $0 \le i \le \pi$,
eccentricity $0 \le e < 1$, eccentric anomaly $0 \le E \le 2 \pi$,
periapsis $0 \le \omega \le \pi$ and mass ratio $M_1/M_2$.  
Other familiar elements (e.g. orbital period $\mathcal{T}$ or orbital speed $\vec{v}_1, \vec{v}_2$)
can be deduced from these seven orbital elements. The elements $x$, $e$,
and $M_A/M_B$ uniquely determine the shape of the both ellipses, while $i$
and $\omega$ determine the orientation of the {\it orbital plane}
(here the $[x,y]$-plane), and a parameter $E$ determines the  
actual position of both components of the binary, ${\cal 1}$ and ${\cal 2}$.  
In case of a symmetric system where $M=M_1=M_2$ and $\vec r_o \equiv \vec r_1 = -\vec r_2$,  
an idealized, circular orbit may be possible.}  
\label{Orbital_Parameters}
\end{figure}

In the original work of Hanbury Brown and Twiss the HBT effect has been exploited in order to 
determine the spatial size of stars.  
Here, in our investigation we will revisit that astronomical problem and apply these new approaches  
to astrophysical situations with the main focus on binary systems. 
Accordingly, each companion of a double star 
is considered to be an individual thermal source of light and the photons of these stars are HBT correlated.  
Using this assumption, we will examine the possible applicability of the 
differential HBT method to determine some orbital parameter of a binary system:  
the semi-major axis $A$, orbital period $\mathcal{T}$, and orbital speed of
the both components $\vec{v}_A,\vec{v}_B$; see Fig.~\ref{Orbital_Parameters}.

At the moment being, there are about $10^5$ binary systems
(resolved, astrometric, eclipsing, spectroscopic) are known,
according to the {\it Washington Double Star Catalog} \cite{WDS}.
However, only for a small part of all these binaries the complete set of
all seven orbital elements have been determined so far.  
Consequently, the reason for our investigation is twofold: 
(i) The suggested approach could demonstrate the presence of the HBT effect in binary systems.  
(ii) Such an approach could allow for determining orbital parameter independently  
from standard astrophysical or astrometric approaches.  

The paper is organized as follows:
In Section \ref{HBT} the HBT effect is reviewed for basic understanding, both in terms
of classical electrodynamics and quantum electrodynamics.  
In Section \ref{Analogy} it is discussed how the advanced HBT approach used in heavy-ion physics has to be modified  
for the case of astrophysics.  
The emission function as essential element of the two-photon correlation function is determined in Section \ref{TEF}.  
Using this modified approach, the two-photon correlation function for a binary system
is determined in Section \ref{PCF}. In Section \ref{SSofcs} the case of one thermal light-source (one star) is 
considered, while the case of two thermal light-sources (binary star) is discussed in Section \ref{FSM}.  
A summary and outlook is given in Section \ref{Summary}.  

Throughout the article, we use Heaviside-Lorentz units: $\epsilon_0 = \mu_0 = c = \hbar = 1$,  
and for the Boltzmann constant $k_B = 1$. Furthermore, the astronomical unit is denoted by  
$1\,{\rm a.u.} = 1.49 \times 10^{11}\,m$. 

\section{HBT effect}\label{HBT}

In 1954, Hanbury Brown and Twiss \cite{HBT1} have realized the idea of an intensity interferometer  
for classical electromagnetic radiation in the radio-band.  
Such intensity interferometry compares intensities of the
incoming radiation field rather than  
amplitudes at different detection points. Originally, the basic
idea was that intensity-measurements are able to improve
the angular resolution of radiation-sources and do not account for
the phase of the electromagnetic fields, hence are insensitive
to phase shifts caused by atmospheric scintillations.
They have first tested their intensity interferometer 
by determining the known  
angular size of the Sun, and later they succeeded in determining the
angular size of the radio sources Cassiopaia A and Cygnus A \cite{HBT1}.

In a first laboratory desk experiment, Hanbury Brown and Twiss 
have extended their new technology into the optical range of 
electromagnetic spectrum. They  
have used a light beam from a mercury vapor lamp, which is actually a
thermal light source, and were able to detect second-order correlations  
among photons \cite{HBT2}, being the discovery of a new physical effect. 
Afterwards, Hanbury Brown and
Twiss succeeded with their technique in determining
the angular size of the star Sirius by detecting
photon correlations in the optical region \cite{HBT3}.

While the correlation of radio waves is an effect which can fully be explained by  
classical electrodynamics, the correlation of photons in the HBT experiment  
initiated a heated debate about the concept of the photon at that time.  
Later, Glauber succeeded with a comprehensive theoretical explanation  
of the HBT effect in a series of articles \cite{Glauber3,Glauber1,Glauber2}.  
Especially, Glauber (i) introduced the concept 
of higher-coherence in quantum electrodynamics \cite{Glauber3},  
(ii) determined the explicit quantumfield-theoretical expression for coherent states   
(Glauber states) \cite{Glauber1}, 
(iii) determined the general expression for the density operator for 
coherent and incoherent states \cite{Glauber2},  
and (iv) has shown  
that the HBT effect does not happen for
Laser light sources (described by Glauber states),
but occurs for extended thermal light
sources (described by incoherent states) \cite{Glauber2}.
These articles were the birth of a new
branch of science, the Quantum optics,
for which he awarded the Physics Nobel Prize in 2005.

In general, in HBT measurements one records photons of a thermal light source by 
two detectors: detector A is located at $\vec{x}_A$ and detects incomming photons  
at time $t_A$, while detector B is located at $\vec{x}_B$ and detects incomming photons 
at time $t_B$. The spatial distance between both detectors is $d_{AB} =  \left| \vec{x}_A - \vec{x}_B\right|$.  
Depending on the concrete experiment under consideration, two specific cases  
have to be distinguished: correlation measurement in space  
($t_A=t_B$ and $\vec{x}_A \neq \vec{x}_B$) and correlation measurement in time  
($t_A \neq t_B$ and $\vec{x}_A = \vec{x}_B$).  
In our investigation, correlation measurements in space are relevant,  
that means the incoming photons are recorded at the same time, $t_A=t_B$,  
while the distance $d_{AB}$ between both detectors is varied. In this way one can  
determine second-order correlations (HBT effect) of incoming photons,  
which are used to determine some physical parameters of the thermal light-source.  
Fundamental aspects of the HBT effect will be demonstrated by means of an elementary model for the thermal radiation. 

\subsection{HBT effect in terms of classical electrodynamics}\label{ED}  

In order to describe the HBT effect it is useful to 
subdivide the surface of a star
with stellar radius $R$
into pointlike regions, each of which is considered as 
an emitter of spherical waves of thermal light, i.e. pointlike sources of 
black-body radiation.
The entire radiation field of the star would finally been 
obtained by a summation over all of these
pointlike regions over the whole surface of the star. 

For the description of spherical waves it is advantageous to 
introduce spherical coordinates $\vec{x} = \left(x,y,z\right) \longrightarrow \vec{r} = \left(r,\theta,\phi\right)$, 
i.e. $x = r\,\cos\theta\,\cos\phi\,,\,y = r\,\cos\theta\,\sin\phi\,,\,z = r\,\cos\theta$. 
Here the origin of coordinate system is, first of all, assumed to 
be located at the center of a pointlike radiation source, hence  
$r = \left|\vec{x}\right|$ being the distance between 
the pointlike region of emission and
some point with spatial coordinate $\vec{x} \leftrightarrow \vec{r}$ (later it will 
be the detector's position). 
In general, the radiative part of electric field 
at sufficiently far distances from a localized 
radiation-source, defined by the wave-zone $k\,r \gg 1$ 
(wave-number $k$ is related to wavelength $\lambda$ in virtue of $k=2\,\pi/\lambda$), is given by 
the following multipole-expansion for one individual frequency $\omega_k = k$ \cite{Jackson}:  
\begin{eqnarray}
\vec{E}\left(\vec{r},t\right) &=& i\,
\frac{{\rm e}^{i\left(k \,r - \omega_k\,t\right)}}{k\,r}  
\sum\limits_{l=1}^{\infty} \sum \limits_{m=-l}^{l} 
\alpha_{l m}\;\vec{g}_{l m}\left(\theta,\phi\right) + c.c.\,,   
\label{Introduction_11}
\end{eqnarray}

\noindent
where we have assumed that the radiation source has no 
magnetic moments; $c.c.$ stands for complex-conjugate and 
$l \ge 1$ because there is no monopole radiation.  
Here, $\vec{g}_{l m} = \left(-i\right)^l\,\vec{n} \times \vec{X}_{l m}$ 
with the unit direction $\vec{n}=\vec{r}/r$ and the  
vector spherical harmonics 
$\vec{X}_{l m} = \vec{l}\,Y_{l m}/\sqrt{l \left(l+1\right)}$, 
the orbital angular-momentum operator is  
$\vec{l} = \vec{r} \times \vec{\nabla}$, 
and the spherical harmonics are $Y_{l m}$. The eigenvalues of the spherical harmonics 
are $\vec{l}^2\,Y_{l m} = l\left(l+1\right) Y_{l m}$ 
and $\vec{l}_z\,Y_{l m}= m\,Y_{l m}$.  
The electric multipole-coefficients are complex-valued numbers,   
$\alpha_{l m} = \left|\alpha_{l m}\right|\,{\rm e}^{i\,\varphi_k}$  with phase $\varphi_k$  
\footnote{The electric multipole-coefficients $\alpha_{l m}$ used in our article are related to the 
electric multipole-coefficients $a_E\left(l,m\right)$ used in \cite{Jackson} via  
$\alpha_{l m} = Z_0\,a_E\left(l,m\right)$, where $Z_0$ is the wave-impedance in vacuum.}.  

In order to get an idea about the magnitude of wave-number k, we 
recall that a star is an almost ideal black body radiator, 
that means the wavelength of maximal intensity of a star is 
determined by Wien's law of displacement,  
which states that the wavelength $\lambda_{\rm max}$ of maximal intensity
of a star with surface temperature $T$ is given by
$\lambda_{\rm max} = 2.9 \times 10^{-3}\,{\rm meter}\,{\rm Kelvin}/T$.
The corresponding wave-vector $k_{\rm max} = 2\,\pi/\lambda_{\rm max}$.
For instance, Sirius has a surface temperature
of about $9900\,{\rm K}$ and we obtain
$\lambda_{\rm max} = 2.92 \times 10^{-7}\,{\rm meter}$ and
$k_{\rm max} = 2.15 \times 10^7\,{\rm meter}^{-1}$.
Therefore, it would be meaningful to determine the HBT
effect with photo-detectors which are sensitive
in the vicinity around $k_{\rm max}$.

\begin{figure}[!ht]
\begin{center}
\includegraphics[scale=0.4]{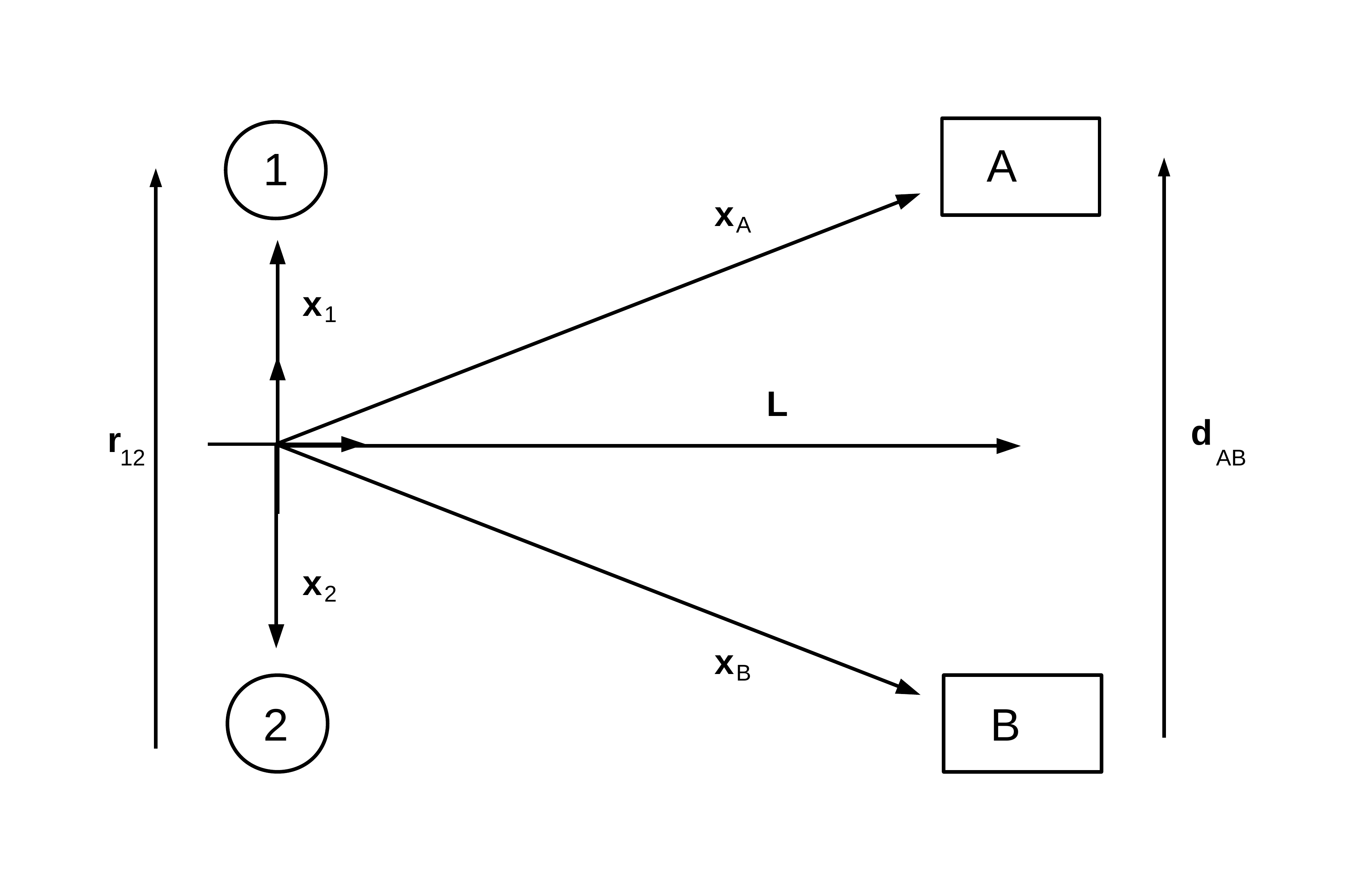}
\end{center}
\caption{Two pointlike regions at $\vec{x}_1$ and $\vec{x}_2$
of the extended light source and
two detectors located at $\vec{x}_A$ and $\vec{x}_B$.
The vectors are defined by
$\vec{r}_{12} = \vec{x}_1 -
\vec{x}_2$, $\vec{d}_{AB} = \vec{x}_A - \vec{x}_B$ and
$\displaystyle \vec{L} = \frac{\vec{x}_A + \vec{x}_B}{2}$.
Note that
$\displaystyle \vec{x}_A = \vec{L} + \frac{\vec{d}_{AB}}{2}$ and
$\displaystyle \vec{x}_B = \vec{L} - \frac{\vec{d}_{AB}}{2}$.}
\label{Electric_Fields}
\end{figure}

To simplify the notation, we assume the photo-detectors to be 
sensitive only for incoming light rays 
which are polarized in an arbitrary but fixed unit-direction 
$\vec{e}$, thus we consider the vector-component of the electric field 
in this direction: 
${\cal E}\left(\vec{r},t\right) = \vec{e} \cdot \vec{E}\left(\vec{r},t\right)$ 
\cite{Glauber3}, and   
we introduce the notation ${\it g}_{l m} = \vec{e} \cdot \vec{g}_{l m}$.  
Actually, the radiation distribution caused by a superposition 
of incoherent multipoles in the pointlike radiation-source
is isotropic \cite{Jackson}, hence independent of $m$. 
However, in HBT experiments one measures individual photons, 
which in the classical electrodynamics correspond 
to one single radiation-mode of the electric 
radiation-field in (\ref{Introduction_11}), given by:  
\begin{eqnarray}
{\cal E}_{l m}\left(\vec{r},t\right) &=& 
i\,\frac{{\rm e}^{i \left(k \,r - \omega_k\,t\right)}}{k\,r}
\alpha_{l m}\;{\it g}_{\,l m}\left(\theta,\phi\right) + c.c.\,.  
\label{Introduction_12}
\end{eqnarray}

\noindent
For HBT effect, it is sufficient to consider the most simple 
case of two specific spherical-wave modes: one mode   
$l_1 m_1$ originates from the pointlike region at $\vec{x}_1$ 
of the star's surface, while another  
mode $l_2 m_2$ originates from the pointlike region at $\vec{x}_2$ 
of the star's surface. The collective electric field  
at detector position $\vec{x}_A$ and at time $t$ is then given 
by a superposition of these waves, cf. Fig.~\ref{Electric_Fields}:  
\begin{eqnarray}
{\cal E}\left(\vec{x}_1,\vec{x}_2,\vec{x}_A,t\right) 
&=& {\cal E}_{l_1 m_1}\left(\vec{x}_1,\vec{x}_A,t\right) +
{\cal E}_{l_2 m_2}\left(\vec{x}_2,\vec{x}_A,t\right).  
\label{Introduction_15}
\end{eqnarray}

\noindent
Using the relation for a spherical wave-mode originating at $\vec{x}_1$ towards detector at $\vec{x}_A$ with momentum  
$\displaystyle \vec{k}_{1A} = k_1 \frac{\vec{x}_A - \vec{x}_1}{\left|\vec{x}_A - \vec{x}_1\right|}$, 
and another spherical wave-mode originating at $\vec{x}_2$ towards detector at $\vec{x}_A$ with momentum  
$\displaystyle \vec{k}_{2A} = k_2 \frac{\vec{x}_A - \vec{x}_2}{\left|\vec{x}_A - \vec{x}_2\right|}$,  
the scalar product takes the form 
\begin{eqnarray}
\vec{k}_{1A} \cdot \left(\vec{x}_A - \vec{x}_1 \right) &=& k_1 \left| \vec{x}_A - \vec{x}_1 \right|\,, 
\label{momentum_1A}
\\
\nonumber\\
\vec{k}_{2A} \cdot \left(\vec{x}_A - \vec{x}_2 \right) &=& k_2 \left| \vec{x}_A - \vec{x}_2 \right|\,, 
\label{momentum_2A}
\end{eqnarray}

\noindent
where the two electromagnetic spherical-waves, originating   
from region $\vec{x}_1$ and $\vec{x}_2$, are given by  
\begin{eqnarray} 
{\cal E}_{l_1 m_1}\left(\vec{x}_1,\vec{x}_A,t\right) 
&=& i\,\left|\alpha_{l_1 m_1}\right|\,{\it g}_{\,l_1 m_1}\,
\frac{{\rm exp} \left[i \left(k_1\,\left| \vec{x}_A - 
\vec{x}_1\right| - \omega_{k_1}\,t + \varphi_{k_1}\right)\right]}
{k_1\;\left| \vec{x}_A - \vec{x}_1\right|} + c.c.\,,  
\label{Introduction_5}
\\
{\cal E}_{l_2 m_2}\left(\vec{x}_2,\vec{x}_A,t\right) 
&=& i\,\left|\alpha_{l_2 m_2}\right|\,{\it g}_{\,l_2 m_2}\,
\frac{{\rm exp} \left[i \left(k_2\,\left|\vec{x}_A - 
\vec{x}_2\right| - \omega_{k_2}\,t + \varphi_{k_2} \right)\right]}
{k_2\;\left|\vec{x}_A - \vec{x}_2\right|} + c.c.\,,  
\label{Introduction_10}
\end{eqnarray} 
 
\noindent
where the phases are written explicitly.
The electric field at detector B is obtained in a very same way, 
where $\vec{x}_A$ is simply replaced by $\vec{x}_B$.  
The correlation function of second-order,  
for these two spherical-waves is defined by  
\ba
C\left(\vec{x}_1,\vec{x}_2,\vec{x}_A,\vec{x}_B\right)
&=& \frac{\langle I\left(\vec{x}_1,\vec{x}_2,\vec{x}_A,t\right)\,
I\left(\vec{x}_1,\vec{x}_2,\vec{x}_B,t\right)\rangle}
{\langle I\left(\vec{x}_1,\vec{x}_2,\vec{x}_A,t\right)\rangle\,
\langle I\left(\vec{x}_1,\vec{x}_2,\vec{x}_B,t\right)\rangle}\,, 
\label{Intensities_12}
\ea

\noindent
where the intensities $I\left(\vec{x}_A,t\right)$ and
$I\left(\vec{x}_B,t\right)$ are defined as square of the absolute
value of electric field and are measured by two
intensity detectors located at $\vec{x}_A$ and
$\vec{x}_B$, 
\begin{eqnarray}
I\left(\vec{x}_1,\vec{x}_2,\vec{x}_A,t\right) &=& 
{\cal E}^{\ast} \left(\vec{x}_1,\vec{x}_2,
\vec{x}_A,t\right)\;{\cal E}\left(\vec{x}_1,
\vec{x}_2,\vec{x}_A,t\right)\,,
\label{Introduction_16}
\\
I\left(\vec{x}_1,\vec{x}_2,\vec{x}_B,t\right) &=& 
{\cal E}^{\ast} \left(\vec{x}_1,\vec{x}_2,
\vec{x}_B,t\right)\;{\cal E}\left(\vec{x}_1,\vec{x}_2,\vec{x}_B,t\right).  
\label{Introduction_17}
\end{eqnarray} 

\noindent
Thermal radiation fields are characterized by the fact that the amplitudes $\alpha_{l m}$ fluctuate independently of one another,
that means their phases $\varphi_k$ are randomly distributed as well as their absolute values $\left|\alpha_{l m}\right|$.  
Accordingly, the thermodynamical average of a function ${\cal O}$ in (\ref{Intensities_12}) implies, first of all, 
an average over the phases and afterwards an average over the absolute values of the amplitudes.  
The average over the phases resembles the thermal average procedure originally introduced by 
the Einstein-Hopf model \cite{Einstein_Hopf}, and is given by:  
\begin{eqnarray}
\langle {\cal O} \rangle &=& \frac{1}{2\pi} 
\int\limits_0^{2 \pi} {\cal O}\left(\varphi\right) \; d \varphi\,. 
\label{Introduction_20}
\end{eqnarray}

\noindent   
By using (\ref{Introduction_15}), (\ref{Introduction_16}) and (\ref{Introduction_17}),  
and performing an average according to (\ref{Introduction_20}) 
one obtains for the correlation function of second-order  
for two spherical-waves with equal frequencies $\omega_{k_1} = \omega_{k_2}$ the following expression:
\ba
&& C\left(\vec{x}_1,\vec{x}_2,\vec{x}_A,\vec{x}_B\right) 
\nonumber\\
\nonumber\\
&& \hspace{-0.5cm} = 1 + 2\,
\frac{\left|\alpha_{l_1 m_1}\right|^2\,\left|\alpha_{l_2 m_2}\right|^2}
{\left(\left|\alpha_{l_1 m_1}\right|^2 + 
\left|\alpha_{l_2 m_2}\right|^2\right)^2}\, 
\cos\bigg[
k \left( \left|\vec{x}_1 - \vec{x}_A \right| + 
\left|\vec{x}_2 - \vec{x}_B\right| - 
\left|\vec{x}_2 - \vec{x}_A \right|
- \left|\vec{x}_1 - \vec{x}_B \right| \right)
\bigg]\,.
\nonumber\\
\label{Introduction_25}
\ea

\noindent
As stated above, we also have to perform a thermal average over the absolute values of the amplitudes in (\ref{Introduction_25}),  
and one easily obtains:  
\footnote{
The square of absolute value of amplitude is proportional 
to the energy: $\left|\alpha_{l m}\right|^2 \sim E_k = \omega_k$. 
Accordingly, the thermal average is determined as follows: 
$\langle\left|\alpha_{l m}\right|^2\rangle_{\rm th}=
 \frac{\int_0^\infty d E_{k} \,E_{k}\,e^{-E_{k}/T}}
           {\int_0^\infty d E_{k}\,e^{-E_{k}/T}}   = T$ 
and 
$\langle\left|\alpha_{l m}\right|^4\rangle_{\rm th}=
\frac{\int_0^\infty d E_{k}\,E_{k}^2\,e^{-\omega_k/T}}
     {\int_0^\infty d E_{k}\,e^{-E_{k}/T}}
= 2\,T^2 = 2\,\langle\left|\alpha_{l m}\right|^2\rangle^2_{\rm th}$.} 
the relations:   
$\langle\left|\alpha_{l_1 m_1}\right|^2 \rangle_{\rm th} = 
\langle\left|\alpha_{l_2 m_2}\right|^2 \rangle_{\rm th}$ and  
$\langle\left|\alpha_{l_1 m_1}\right|^2\;
\left|\alpha_{l_2 m_2}\right|^2\rangle_{\rm th} 
= 2\,\langle\left|\alpha_{l_1 m_1}\right|^2\rangle^2_{\rm th}$.  
Then, the correlator in (\ref{Introduction_25}) finally yields 
\cite{Scully}:
\ba
C\left(\vec{x}_1,\vec{x}_2,\vec{x}_A,\vec{x}_B\right) &=&  
1 + \cos\bigg[
k \left( \left|\vec{x}_1 - \vec{x}_A \right| + \left|\vec{x}_2 - 
\vec{x}_B\right| - \left|\vec{x}_2 - \vec{x}_A \right|
- \left|\vec{x}_1 - \vec{x}_B \right| \right)
\bigg]\,.
\nonumber\\
\label{Introduction_26}
\ea

\noindent
There are two important limits for the correlation 
function in (\ref{Introduction_26}), namely 
the case of stars and the case of heavy-ion collisions, 
cf. Refs. \cite{Humanic,Heinz_Jacak}. 
Both limiting cases we shall consider in more detail.

\subsubsection{Correlation function in the limit of stars}

In case of a star we have the following limits: $L \gg r_{12} \gg d_{AB}$,
where $L$ is the distance between star and detectors,
$r_{12}$ is the distance among two pointlike regions on the star's surface,
and $d_{AB}$ is the distance between both detectors, 
see Fig.\ref{Electric_Fields}.
In these limits, the argument of the cosine-function 
in (\ref{Introduction_26}) can
considerably been simplified and we obtain for stars (for more details see Section \ref{PCF}):  
\ba
C_{\rm Star}
\left(\vec{x}_1,\vec{x}_2,\vec{x}_A,\vec{x}_B\right) &=&
1 + \cos \left(k\,\frac{\vec{d}_{AB}\cdot\vec{r}_{12}}{L}\right),
\label{Introduction_27}
\ea

\noindent
where $k = 2\,\pi/\lambda$ is the wave-vector of the 
radiation emitted by the pointlike sources, 
$\vec{d}_{AB} = \vec{x}_A - \vec{x}_B$ is the vector from
detector A to detector B and 
$\vec{r}_{12} = \vec{x}_1 - \vec{x}_2$ is the vector from source-point 
$\vec{x}_1$ to the source-point $\vec{x}_2$ on the star's surface. 

So far we have considered two pointlike thermal light-sources of the star.  
The second-order correlation function of classical thermal radiation emitted 
by the entire surface of a star is written as follows:
\ba
C\left(\vec{x}_A,\vec{x}_B\right) &=&
\frac{\langle I\left(\vec{x}_A,t\right)\;I\left(\vec{x}_B,t\right)\rangle}
{\langle I\left(\vec{x}_A,t\right)\rangle\;
\langle I\left(\vec{x}_B,t\right)\rangle}\;. 
\label{Correlation_Function1}
\ea

\begin{figure}[!ht]
\begin{center}
\includegraphics[scale=0.4]{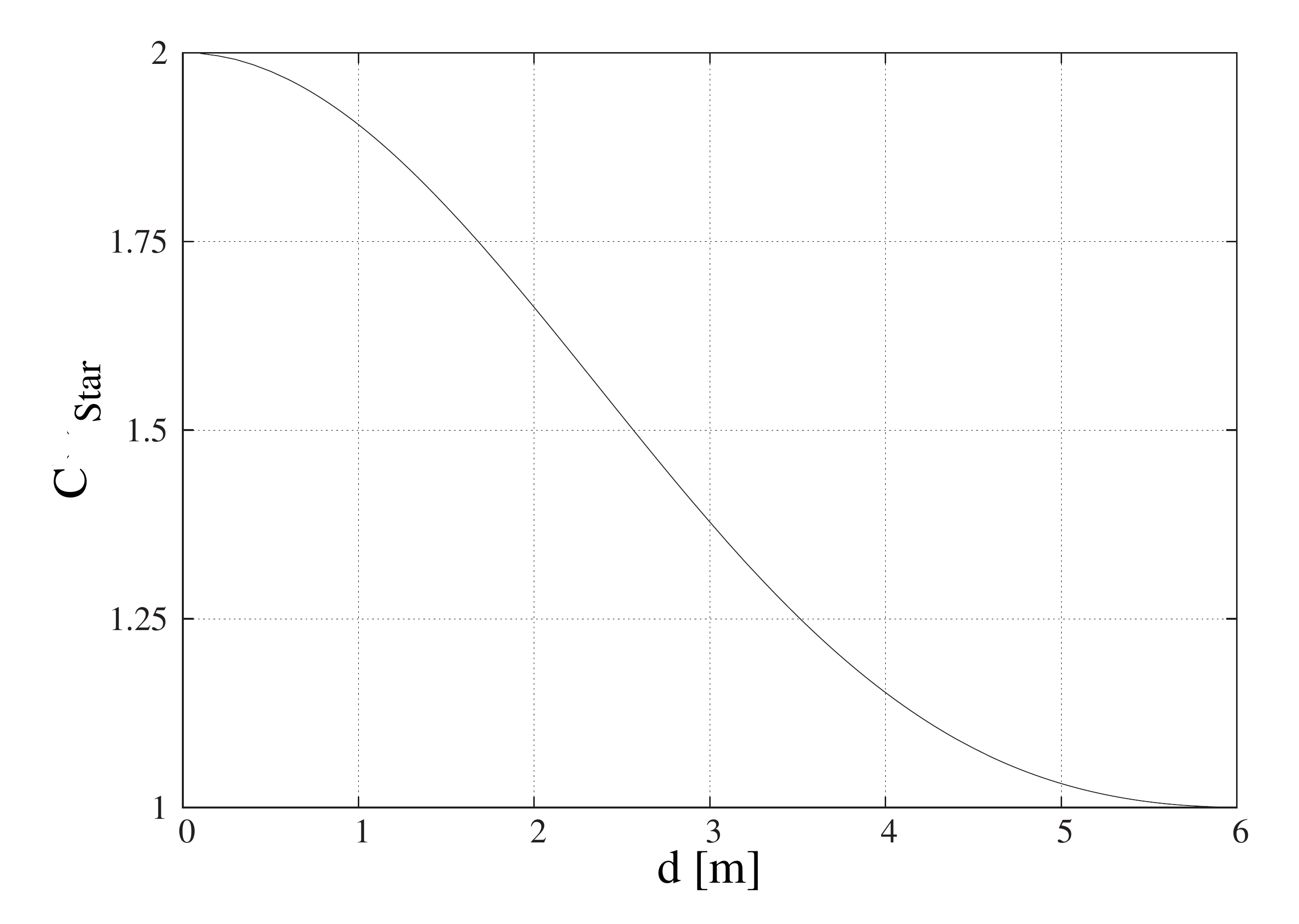}
\end{center}
\vskip -4mm
\caption{The normalized correlation function $C_{\rm Star}$ as given by Eq.~(\ref{Introduction_35b})  
for the following parameter (Sirius): stellar radius $R = 1.2 \times 10^{9}\,{\rm m}$,  
distance between star and detectors $L = 8.6 \,{\rm lightyear} = 8.136 \times 10^{16}\,{\rm m}$,  
and wave-vector $k=2.15 \times 10^7\,{\rm m}^{-1}$. By comparing with (\ref{Introduction_36}) we deduce an opening angle  
$\Theta = 2\,R/L = 2.9 \times 10^{-8}\,{\rm rad}$.}
\label{Cor_Function-expl}
\end{figure}

\noindent
The opacity of a star (impenetrability of electromagnetic radiation or visible light)  
is very high so that the radiation field originates from the star's surface only, more accurately    
the light is emitted by the photosphere of the star. Accordingly, we can obtain the correlation function  
in (\ref{Correlation_Function1}) by a summation of the expression (\ref{Introduction_27})  
over all possible configurations of regions $\vec{x}_1$ and $\vec{x}_2$ of the entire surface of a star.  
Consequently, we have to integrate over the surface of a star facing the observer ($A_{\rm star} = \pi\,R^2$) as follows: 
\begin{eqnarray}
C_{\rm Star}\left(\vec{x}_A,\vec{x}_B\right) 
&=& \frac{1}{A_{\rm star}^2} \int_{A_{\rm star}} d^2 x_1 \;
\int_{A_{\rm star}} d^2 x_2\;
C\left(\vec{x}_1,\vec{x}_2,\vec{x}_A,\vec{x}_B\right)
\label{Introduction_35a}
\\
\nonumber\\
&=& 1 + \frac{4\,L^2}{k^2\,d_{AB}^2\,R^2}\,J_1^2 \left(\frac{k\,R\,d_{AB}}{L}\right),  
\label{Introduction_35b}
\end{eqnarray}

\noindent
where $J_1$ is the Bessel function of first kind.
The result in (\ref{Introduction_35b}) reflects the Siegert relation for thermal light in classical electrodynamics \cite{Siegert}, 
which relates second-order and first-oder correlation function.  
An example for the correlation function in (\ref{Introduction_35b}) is
given in Fig.~\ref{Cor_Function-expl} for Sirius, repeating in this way the intensity interferometry analysis of
Hanbury Brown and Twiss \cite{HBT3}.

In HBT experiments one determines the angular size of the star, $\Theta = 2\,R/L$,  
as seen by an observer at distance $L$, so that the correlator in (\ref{Introduction_35b}) reads  
\begin{eqnarray}
C_{\rm Star}\left(\vec{x}_A,\vec{x}_B\right) &=&
1 + \frac{16}{k^2\,d_{AB}^2\,\Theta^2}\,J_1^2 \left(\frac{k\,d_{AB}\,\Theta}{2}\right),
\label{Introduction_36}
\end{eqnarray}
 
\noindent
which agrees, for instance, with the correlation-function given by Eq.~(2) in \cite{ChenEtal2011}; 
recall $k=2\,\pi/\lambda$. 
The correlator (\ref{Introduction_36}) varies as a function of the telescope separation $d_{AB}$.
Thus, by varying the separation of the detectors $d_{AB}$, one may deduce
the apparent angle $\Theta$ of a remote star, even if the source is optically not resolvable.
In real experiments one compares the correlator determined by real measurements with the
correlator in (\ref{Introduction_36}) (fitting procedure).

\subsubsection{Correlation function in the limit of heavy-ion collisions}

In case of heavy-ion collisions (HIC) we have the following limits:
$L \gg d_{AB} \gg r_{12}$,
where $L$ is the distance between fireball and detectors,
$r_{12}$ is the distance among two pointlike regions on the fireball's 
surface
and $d_{AB}$ is the distance between both detectors, 
see Fig.\ref{Electric_Fields}.
In these limits, the cosine-function in (\ref{Introduction_26})
can be considerably simplified and is given by
\ba
C_{\rm HIC}
\left(\vec{x}_1,\vec{x}_2,\vec{x}_A,\vec{x}_B\right)  &=&
1 + \cos \left(k\,\frac{\vec{d}_{AB}\cdot\vec{r}_{12}}{L}\right).  
\label{wave_function_HIC_14}
\ea

\noindent
For comparison with the heavy-ion collision results we recall the 
practical relation $\vec q = \vec q_{AB} = k\, \vec d_{AB} / L$,
because there  the distance of the detectors, $d_{AB}$,  is not measured 
directly.
We recognize that the correlation function in the limits of heavy-ion 
collisions in (\ref{wave_function_HIC_14}) agrees with the 
correlation function in the limit for stars (\ref{Introduction_27}).  
Also in case of heavy-ion collisions one has to 
sum over all individual pointlike
regions over the entire fireball created in the 
heavy-ion collision process. However,
the description of HBT in heavy-ion collisions is 
more complicated than for stars, because
the fireball expands rapidly in time and space, 
and therefore a simple two-dimensional integration procedure
like in (\ref{Introduction_35a}) is inapplicable, 
instead a four-dimensional description is necessary.

\subsection{HBT effect in terms of quantum electrodynamics}\label{QED}  

The surprising discovery of quantum correlations of 
second-order in ordinary light by
Hanbury Brown and Twiss  came to pass when they applied 
photodetectors  in the domain of the optical spectrum \cite{HBT2,HBT3}.
The theoretical description of a localized absorption process of a
photon by photodetectors (e.g. photodiodes, 
photomultipliers, CCD's) implies the electromagnetic field to 
be considered as made of by photons instead of 
classical waves. One aspect thereof is the symmetrization 
of the photon wavefunction under coordinate-exchange of two 
indistinguishable photons,  
a term which, however, does not occur in classical electrodynamics.  
Accordingly, the use of a photon-counter implies that the
electric radiation-field
$\vec{E}\left(\vec{r},t\right)$
in (\ref{Introduction_11}) must have to be described as a
second-quantized quantum-field, indicated by a hat: $\hat{\vec E}\left(\vec{r},t\right)$ 
\cite{Akhiezer_Berestezki,Davydov}:
\begin{eqnarray}
\hat{\vec E}\left(\vec{r},t\right) &=&
\hat{\vec E}^{\left(+\right)}\left(\vec{r},t\right)
+ \hat{\vec E}^{\left(-\right)}\left(\vec{r},t\right),
\label{Electric_Field_QFT}
\\
\nonumber\\
\hat{\vec E}^{\left(+\right)}\left(\vec{r},t\right) &=&
+ i\,\frac{{\rm e}^{+ i
\left(k \,r - \omega_k\,t\right)}}{k\,r}
\sum\limits_{l=1}^{\infty} \sum \limits_{m=-l}^{l}
\hat{a}_{l m}\;\vec{g}_{l m}\left(\theta,\phi\right),
\label{Electric_Field_QFT_Plus}
\\
\nonumber\\
\hat{\vec E}^{\left(-\right)}\left(\vec{r},t\right) &=&
- i\,\frac{{\rm e}^{- i \left(k \,r -
\omega_k\,t\right)}}{k\,r}
\sum\limits_{l=1}^{\infty} \sum \limits_{m=-l}^{l}
\hat{a}^{\dagger}_{l m}\;\vec{g}^{\ast}_{l m}\left(\theta,\phi\right).
\label{Electric_Field_QFT_Minus}
\end{eqnarray}

\noindent
By comparing the second-quantized electric field operator in
(\ref{Electric_Field_QFT}) with the classical
electric field in (\ref{Introduction_11}), one recognizes that there is
formally a replacement of the classical amplitudes by operators  
\footnote{A more accurate second-quantization procedure is described in detail in \cite{Akhiezer_Berestezki,Davydov} and takes 
into account the spin of the photons ($s_z=-1,0,+1$) by  
introducing the total momentum-operator $\vec{j} = \vec{l} + \vec{s}$, where $\vec{s}$ is an operator acting in the Hilbert-space  
of spin-states of photons. However, the exlicit notation of a spin quantum-number would not change the basic arguments given.}:  
$\alpha_{l m} \rightarrow \hat{a}_{l m}$
and $\alpha^{\ast}_{l m} \rightarrow \hat{a}^{\dagger}_{l m}$, where
$\hat{a}_{l m}^{\dagger}$ is the Hermitian
adjoint of $\hat{a}_{l m}$.
These photon-operators act in the Fock-space of
photon states and obey commutator relations in angular-momentum space,
given by \cite{Akhiezer_Berestezki,Davydov}:
\ba
\left[\hat{a}_{l_1 m_1}\,,\,\hat{a}^{\dagger}_{l_2 m_2} \right]_{-} &=&
\omega_k\,\delta_{l_1 l_2}\,\delta_{m_1 m_2}\,, \quad
\left[\hat{a}_{l_1 m_1}\,,\,\hat{a}_{l_2 m_2} \right]_{-} = 0\,, \quad
\left[\hat{a}^{\dagger}_{l_1 m_1}\,,\,\hat{a}^{\dagger}_{l_2 m_2} \right]_{-} = 0\,.
\label{Commutator_1}
\ea

\noindent
The operator $\hat{a}^{\dagger}_{l_1 m_1}$ creates
one photon with angular-momentum $l_1 m_1$, while
the operator $\hat{a}_{l_1 m_1}$ annihilates one photon with
angular-momentum $l_1 m_1$.
In particular, the vacuum state is defined by $\hat{a}_{l_1 m_1} |
{\rm vac}\rangle = 0$, and
$| l_1 m_1 \rangle = \hat{a}_{l_1 m_1}^{\dagger} |
{\rm vac} \rangle$ represents a one-photon state,
while $\displaystyle | l_1 m_1\; l_2 m_2 \rangle =
\frac{1}{\sqrt{2}}\,\hat{a}_{l_1 m_1}^{\dagger}
\hat{a}_{l_2 m_2}^{\dagger} | {\rm vac} \rangle$
represents a two-photon state in the angular-momentum space.
Accordingly, the positive-energy part (\ref{Electric_Field_QFT_Plus})
annihilates one photon localized at $r,\theta,\phi,t$
in coordinate-space, while the negative-energy part
(\ref{Electric_Field_QFT_Minus}) creates one photon
localized at $r,\theta,\phi,t$ in coordinate-space.

Like in the classical case, we assume the detectors are
sensitive only to those photons which are polarized in an arbitrary
but fixed unit-direction $\vec{e}$,
and consider the vector-component of the electric
field-operator in this direction:
$\hat{\cal E}\left(\vec{r},t\right) = \vec{e}
\cdot \hat{\vec E}\left(\vec{r},t\right)$ \cite{Glauber3}.
One individual mode of the electric field-operator is then given by
\ba
\hat{\cal E}\left(\vec{r},t\right) &=&
\hat{\cal E}^{\left(+\right)}\left(\vec{r},t\right) +
\hat{\cal E}^{\left(-\right)}\left(\vec{r},t\right)\,,
\label{Electric_field_operator}
\\
\nonumber\\
\hat{\cal E}^{\left(+\right)}\left(\vec{r},t\right) &=&
+ i\,\frac{{\rm e}^{+ i \left(k \,r -
\omega_k\,t\right)}}{k\,r}
\hat{a}_{l m}\;{\it g}_{\,l m}\left(\theta,\phi\right),
\label{Electric_field_operator_positive}
\\
\nonumber\\
\hat{\cal E}^{\left(-\right)}\left(\vec{r},t\right) &=&
- i\,\frac{{\rm e}^{- i \left(k \,r -
\omega_k\,t\right)}}{k\,r}
\hat{a}^{\dagger}_{l m}\;{\it g}^{\ast}_{\,l m}
\left(\theta,\phi\right).
\label{Electric_field_operator_negative}
\ea

\noindent
The field-operator in Eq.~(\ref{Electric_field_operator})
is the quantum-field analogon
of the classical electric field in Eq.~(\ref{Introduction_12}).

The theory of HBT effect in terms of quantum
electrodynamics has been worked out
by Glauber in a series of articles \cite{Glauber3,Glauber1,Glauber2}.
We shall recall some parts of this theory which
are relevant for our investigation.

The initial state before a one-photon detection at time $t$
is denoted by $|i\rangle$, and the final state after detection is
denoted by $|f\rangle$ which contains one photon less because
it has been absorbed by the detector located at $\vec{x}_A$.
The final states form a complete set of states,
$\sum_f | f\rangle\,\langle f| = {\bf 1}\hspace{-4pt}1$.
The corresponding amplitude for a one-photon absorption is
given by the following matrix element:
$\langle f |
\hat{\cal E}^{\left(+\right)}\left(\vec{x}_A,t\right)
| i \rangle$.
The probability of this process is given by the square of the
absolute value of this amplitude (Fermi's golden rule),
$
W^{\left(1\right)}_{if} =
\left|\langle f | \hat{\cal E}^{\left(+\right)}\left(\vec{x}_A,t\right)
| i \rangle\right|^2
$.
If one wants to determine the total probability that one
photon is absorbed at time $t$ by an ideal photon detector located
at $\vec{x}_A$ and irrespective of the final state,
then one has to sum over all final states and obtains,
cf. Eq.~(2.15) in \cite{Glauber3}:
\ba
W^{\left(1\right)}_i\left(\vec{x}_A,t\right) &=&
\sum\limits_f
\left|\langle f |
\hat{\cal E}^{\left(+\right)} \left(\vec{x}_A,t\right)
| i \rangle\right|^2
= \langle i | \hat{\cal E}^{\left(-\right)}\left(\vec{x}_A,t\right)
\hat{\cal E}^{\left(+\right)}\left(\vec{x}_A,t\right) | i \rangle\,,
\label{Probability_1}
\ea
\noindent
where in the last term a summation over the
complete set of final states has been performed.
If the initial state $|i\rangle$ is not a pure state but a statistical
mixture of states (thermal bath of photons),
then the initial state is described by a statistical
operator $\hat{\rho}$ and the average in (\ref{Probability_1})
has to be determined by the trace over the density operator:
$
\langle i | \hat{\cal O} | i \rangle \rightarrow {\rm Tr}
\left(\hat{\rho}\,\hat{\cal O}\right)
$.
Similarly, in case of two-photon detection at time $t$ by two
detectors located at $\vec{x}_A$ and $\vec{x}_B$,
the amplitude of this process reads
$
\langle f |
\hat{\cal E}^{\left(+\right)}\left(\vec{x}_B,t\right)\,
\hat{\cal E}^{\left(+\right)}\left(\vec{x}_A,t\right)
| i \rangle ,
$
and the probability of this process is (Fermi's golden rule)
$
W^{\left(2\right)}_{if} =
\left|\langle f |
\hat{\cal E}^{\left(+\right)}\left(\vec{x}_A,t\right)
\hat{\cal E}^{\left(+\right)}\left(\vec{x}_B,t\right)
| i \rangle\right|^2
$.
The total probability, that two photons are absorbed at
time $t$ by two ideal photon detectors located
at $\vec{x}_A$ and $\vec{x}_B$, and irrespective of the final state
is given by, cf. Eq.~(2.17) in \cite{Glauber3}:
\ba
W^{\left(2\right)}_{i}\left(\vec{x}_A, \vec{x}_B\right) &=&
\sum\limits_f \left| \langle f |
\hat{\cal E}^{\left(+\right)}\left(\vec{x}_B,t\right)\,
\hat{\cal E}^{\left(+\right)}\left(\vec{x}_A,t\right)
| i \rangle \right|^2
\nonumber\\
\nonumber\\
&=& \langle i |
\hat{\cal E}^{\left(-\right)}\left(\vec{x}_A,t\right)\,
\hat{\cal E}^{\left(-\right)}\left(\vec{x}_B,t\right)\,
\hat{\cal E}^{\left(+\right)}\left(\vec{x}_B,t\right)\,
\hat{\cal E}^{\left(+\right)}\left(\vec{x}_A,t\right) | i \rangle\,,
\label{Probability_2}
\ea

\noindent
where in the last line a summation over the complete set of
final states has been performed.
The expression in (\ref{Probability_2}) describes basically the
correlation of a two-photon absorption process, namely one photon
is absorbed at detector position $\vec{x}_A$ and at the same instant of
time another photon is absorbed at detector position $\vec{x}_B$.
In this sense one can call this expression correlation function. For
a theoretical description of real correlation experiments
it is convenient to consider, first of all, two pintlike regions
located at $\vec{x}_1$
and $\vec{x}_2$ on the star's surface, each of which emits one
spherical photon-mode.
Then, we arrive at the following correlation function,
\ba
C\left(\vec{x}_1, \vec{x}_2,
\vec{x}_A, \vec{x}_B\right) &\!\!=\!\!&
\frac{\sum\limits_f \left| \langle f |
\hat{\cal E}^{\left(+\right)}\left(\vec{x}_1,\vec{x}_2,\vec{x}_B\right)\,
\hat{\cal E}^{\left(+\right)}\left(\vec{x}_1,\vec{x}_2,\vec{x}_A\right)
| i \rangle \right|^2}
{\sum\limits_f \left| \langle f |
\hat{\cal E}^{\left(+\right)}\left(\vec{x}_1,\vec{x}_2,\vec{x}_B\right)
| i \rangle \right|^2
\sum\limits_f \left| \langle f |
\hat{\cal E}^{\left(+\right)}\left(\vec{x}_1,\vec{x}_2,\vec{x}_A\right)
| i \rangle \right|^2} \,,
\label{Correlation_Function2}
\ea
\noindent
where we also have introduced a normalized form of
(\ref{Probability_2}), cf. Eq.~(4.3) in \cite{Glauber3}.
The correlation function (\ref{Correlation_Function2})
determines the rate of coincidences
in the photon count rate of two detectors at
$\vec{x}_A$ and $\vec{x}_B$, and is the
quantum-field theoretical analog of the classical
correlator (\ref{Intensities_12}) for
two pointlike regions of thermal radiation.
In order to determine the correlation function in
(\ref{Correlation_Function2}),
the electric field-operators in (\ref{Correlation_Function2})
can been obtained by
a second-quantization procedure of the classical expressions
in (\ref{Introduction_15}) - (\ref{Introduction_10})
with the aid of formal replacements of the classical
amplitudes by operators 
$\alpha_{l m} \rightarrow \hat{a}_{l m}$
and $\alpha^{\ast}_{l m} \rightarrow \hat{a}^{\dagger}_{l m}$,
\begin{eqnarray}
\hat{\cal E}^{\left(+\right)}\left(\vec{x}_1,\vec{x}_2,\vec{x}_A,t\right) &=&
\hat{\cal E}^{\left(+\right)}\left(\vec{x}_1,\vec{x}_A,t\right) +
\hat{\cal E}^{\left(+\right)}\left(\vec{x}_2,\vec{x}_A,t\right),
\label{Electric_Field_Operator_A}
\end{eqnarray}

\noindent
where these two positive-energy field-operators
\begin{eqnarray}
\hat{\cal E}^{\left(+\right)}\left(\vec{x}_1,\vec{x}_A,t\right) &=&
+ i\,\hat{a}_{l_1 m_1}\,{\it g}_{\,l_1 m_1}\,
\frac{{\rm exp} \left[+ i \left(k_1\,\left| \vec{x}_A - \vec{x}_1\right| -
\omega_{k_1}\,t\right)\right]}
{k_1\;\left| \vec{x}_A - \vec{x}_1\right|}\,,
\label{Electric_Field_Operator_B}
\\
\hat{\cal E}^{\left(+\right)}\left(\vec{x}_2,\vec{x}_A,t\right) &=&
+ i\,\hat{a}_{l_2 m_2}\,{\it g}_{\,l_2 m_2}\,
\frac{{\rm exp} \left[+ i \left(k_2\,\left|\vec{x}_A - \vec{x}_2\right| -
\omega_{k_2}\,t\right)\right]}
{k_2\;\left|\vec{x}_A - \vec{x}_2\right|}\,.
\label{Electric_Field_Operator_C}
\end{eqnarray}

\noindent
The field-operator at detector positon $\vec{x}_B$ can be
obtained by a replacement
of $\vec{x}_A$ by $\vec{x}_B$ in Eqs.~(\ref{Electric_Field_Operator_A}) -
(\ref{Electric_Field_Operator_C}).
In the initial state in the nominator and denominator in
(\ref{Correlation_Function2}) there are two photons:
$\displaystyle | i \rangle = \frac{1}{\sqrt{2}} |l_1 m_1 \; l_2 m_2 \rangle $.
In the nominator in (\ref{Correlation_Function2}) the final
state is the vacuum state
$\langle f | = \langle {\rm vac} |$, while in the
denominator in (\ref{Correlation_Function2}) the final state
is a one-photon state, either $\langle f | = \langle l_1 m_1 |$ or
$\langle f | = \langle l_2 m_2 |$, depending on which photon
has not been detected.
We assume that both emitted photons have the same energy $k_1 = k_2$
but, for simplicity,
here we take either $l_1 \neq l_2$ or $m_1 \neq m_2$.
Then, by inserting the field-operators (\ref{Electric_Field_Operator_A})
into (\ref{Correlation_Function2}) we obtain the following expression
for the correlation function:
\ba
C
\left(\vec{x}_1,\vec{x}_2,\vec{x}_A, \vec{x}_B\right)
&=& \left|\Psi_{12}\right|^2 \,,
\label{Correlation_Function_12_C}
\ea

\noindent
where the photon wavefunction
\footnote{Let us briefly make some comments about this notion. According
to a strict mathematical statement of {\it Landau} and {\it Peierls}
\cite{Landau_Peierls}, there is actually no photon wavefunction
in quantum-field theory. On the other side, interference effects in
double-slit
experiments have shown that the probability-density
for photons have some kind of analogy with electrons,
e.g. \cite{Photon_WF1},
and have allowed to reintroduce the concept of a
photon wavefunction in an appropriate meaning, for a review we refer to
\cite{Photon_WF2,Photon_WF3}.}
is given by
\ba
\Psi_{12} \left(\vec{x}_1, \vec{x}_2, \vec{x}_A,\vec{x}_B\right) &=&
\frac{1}{\sqrt{2}} \left[
{\rm e}^{i\,k\,\left|\mbox{\scriptsize\boldmath $x_1$} -
\mbox{\scriptsize\boldmath $x_A$}\right|}\;
{\rm e}^{i\,k\,\left|\mbox{\scriptsize\boldmath $x_2$} -
\mbox{\scriptsize\boldmath $x_B$}\right|}
+ {\rm e}^{i\,k\,\left|\mbox{\scriptsize\boldmath $x_2$} -
\mbox{\scriptsize\boldmath $x_A$}\right|}\;
{\rm e}^{i\,k\,\left|\mbox{\scriptsize\boldmath $x_1$} -
\mbox{\scriptsize\boldmath $x_B$}\right|}
\right]\,.
\label{WF_5}
\ea
This wavefunction is valid both for heavy-ion collisions and
stars, and symmetric under exchange
of source-coordinates $\vec{x}_1 \leftrightarrow \vec{x}_2$ and
detector-coordinates $\vec{x}_A \leftrightarrow \vec{x}_B$.
After an experimental measurement the wavefunction collapses and
disappears completely.
Let us also recall, that $\Psi_{12}$ is not an observable,
while $\left|\Psi_{12}\right|^2$ can be determined by experiments.
For the absolute square of photon wavefunction (\ref{WF_5}) we obtain
\ba
\left|\Psi_{12}\right|^2 &=& 1 + \cos\bigg[
k \Big( \left|\vec{x}_1 - \vec{x}_A \right| + \left|\vec{x}_2 -
        \vec{x}_B\right| - \left|\vec{x}_2 - \vec{x}_A \right|
- \left|\vec{x}_1 - \vec{x}_B \right| \Big)
\bigg]\,.
\label{WF_1}
\ea

\noindent
We recognize, that the absolute square of wavefunction
in (\ref{WF_1}) equals the classical
correlation function given by (\ref{Introduction_26}).

\begin{figure}[!ht]
\begin{center}
\includegraphics[scale=0.45]{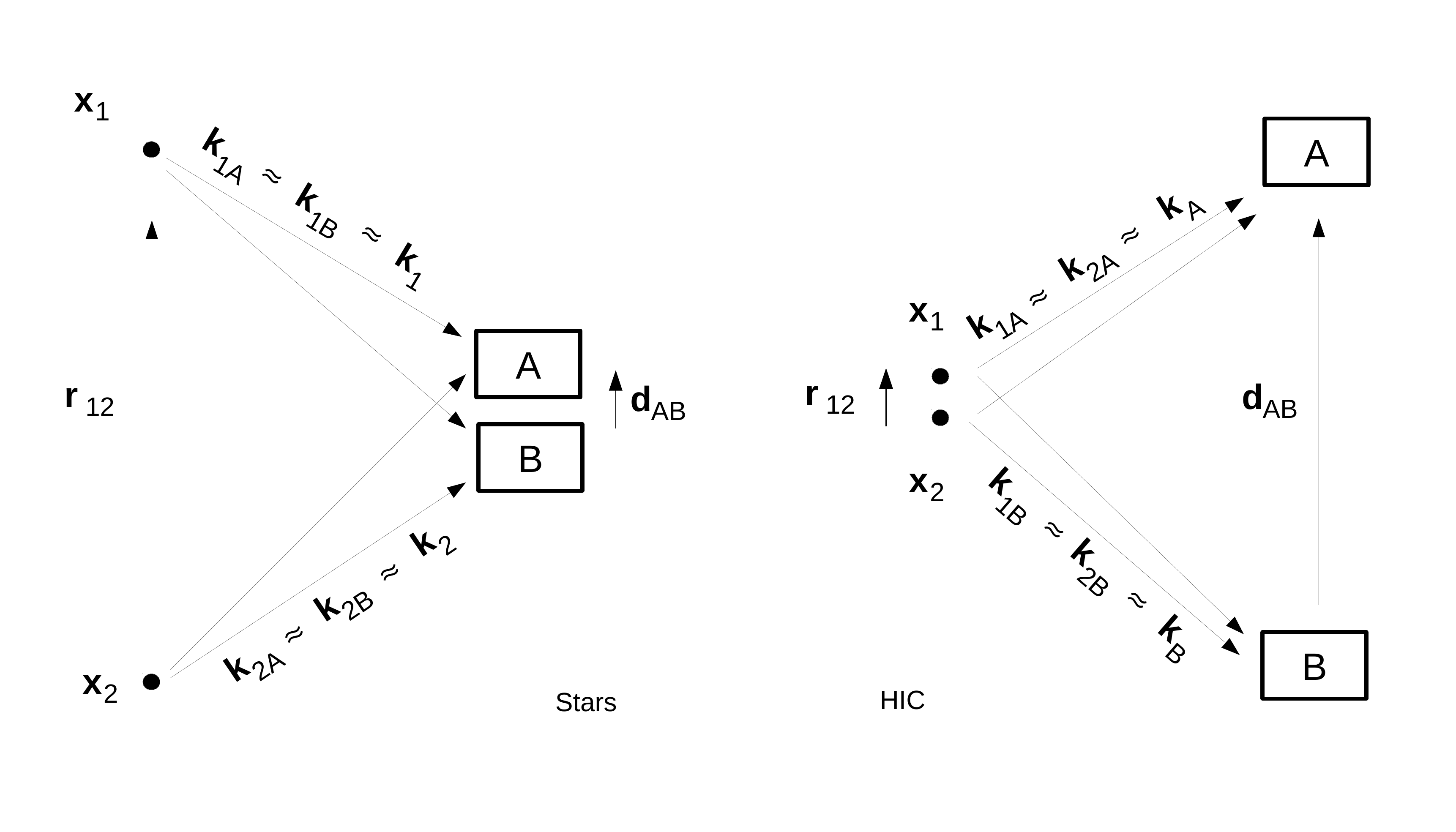}
\end{center}
\vskip -4mm
\caption{The extreme limits for the momenta. Left pannel: 
photon-momenta in case of stars $r_{12} \gg d_{AB}$. 
In this limit the wavefunction $\Psi_{12}$ in (\ref{WF_5}) simplifies to $\Psi_{12}^{\rm Star}$ in (\ref{wave_function_stars_5}). 
Right pannel: photon-momenta in case of heavy-ion collisions $r_{12} \ll d_{AB}$. 
In this limit the wavefunction $\Psi_{12}$ in (\ref{WF_5}) simplifies to $\Psi_{12}^{\rm HIC}$ in (\ref{wave_function_HIC_5}).}  
\label{Wavefunction}
\end{figure}

The symmetrization of wavefunction in (\ref{WF_5}) is a fundamental law
of quantum mechanics and is independent of any specific
performance of an experiment. In other words, the photons,
during their propagation from the star (or fireball of a
heavy-ion collision process) to the
detectors, do not take care about the existence of some
detectors anywhere in the universe.
For instance, in real HBT experiments in heavy-ion collisions
one might want to
install about $3000$ detectors around the fireball
in order to minimize the loss of photons which have been emitted by
the fireball and to improve the statistics and accuracy.
Finally,  
one is searching in the experimental
data for all two-photon correlations (HBT effect) measured
by any two detectors among these $3000$ detectors. But the 
photon (or pion) wavefunction is symmetric
in the coordinates of the fireball as well as in the coordinates of
the final two detectors. The same symmetry exists in case of 
HBT measurements for stars.  
However, different experiments have different geometrical configurations.
Depending on these specific experimental configurations, see Fig.~\ref{Wavefunction}, some parts of the
wavefunction in (\ref{WF_5}) become important, while other parts 
in the wavefunction become
negligible. For instance, in case of stars the wave-function is 
given by Eq.~(1) in \cite{Humanic}, 
while in case of heavy-ion collisions the wave-function is 
given by Eq.~(8) in \cite{Humanic}.  

Actually, more important is the absolute square of the wave-function.  
Especially, like in the classical theory, there are two relevant 
limits for the absolute value in Eq.~(\ref{WF_1}), stars and heavy-ion collisions, which will be considered 
in the following.

\subsubsection{Wavefunction for stars}\label{WF_Star}

In case of a star we have the following limits; see left pannel of Fig.~\ref{Wavefunction}: 
$L \gg r_{12} \gg d_{AB}$, where $L$ is the distance between star and detectors,  
$r_{12}$ is the distance among two pointlike regions on the star's surface, and $d_{AB}$ is the distance between both detectors,  
see Fig.\ref{Electric_Fields}. In these limits, photon wavefunction in (\ref{WF_5}) can considerably be simplified and is given by;  
cf.~\cite{Humanic}:  
\ba
\Psi_{12}^{\rm Star} &=& \frac{1}{\sqrt{2}}
\bigg[
{\rm e}^{i\left[\mbox{\scriptsize\boldmath$k_1$}\cdot\left(\mbox{\scriptsize\boldmath$x_A$}-\mbox{\scriptsize\boldmath$x_1$}\right) 
+ \mbox{\scriptsize\boldmath$k_2$}\cdot\left(\mbox{\scriptsize\boldmath $x_B$}-\mbox{\scriptsize\boldmath $x_2$}\right)\right]} 
+ {\rm e}^{i\left[\mbox{\scriptsize\boldmath$k_1$}\cdot\left(\mbox{\scriptsize\boldmath$x_B$}-\mbox{\scriptsize\boldmath$x_1$}\right)
+ \mbox{\scriptsize\boldmath$k_2$}\cdot\left(\mbox{\scriptsize\boldmath $x_A$}-\mbox{\scriptsize\boldmath $x_2$}\right)\right]}
\bigg]\,,
\label{wave_function_stars_5}
\ea

\noindent
where the wave-vector $\vec{k}_1$ is directed from spatial coordinate $\vec{x}_1$ of the star's surface to detector A or B,  
and wave-vector $\vec{k}_2$ is directed from spatial coordinate $\vec{x}_2$ of the star's surface to detector A or B.  
The derivation of Eq.~(\ref{wave_function_stars_5}) from Eq.~(\ref{WF_5}) can be performed like in Section \ref{PCF}.  
The square of absolute value of this wavefunction is given by   
\ba
\left|\Psi_{12}^{\rm Star}\right|^2 &=&
1 + \cos \left(k\,\frac{\vec{d}_{AB}\cdot\vec{r}_{12}}{L}\right),
\label{wave_function_stars_15}
\ea

\noindent
where $\vec{d}_{AB} = \vec{x}_B - \vec{x}_A$ is the vector from detector A towards detector B, 
and $\vec{r}_{12} = \vec{x}_2 - \vec{x}_1$ is the vector from surface-point $\vec{x}_1$ towards  
the surface-point $\vec{x}_2$ of the star, $k= \left|\vec{k}_1\right| = \left|\vec{k}_2\right|$, and 
we have used the relation $\displaystyle \vec{k}_2-\vec{k}_1=k\,\frac{\vec{r}_{12}}{L}$ which is valid for the geometry of stars; 
see left pannel in Fig.~\ref{Wavefunction}. 
Obviously, (\ref{wave_function_stars_15}) equals the classical expression obtained in (\ref{Introduction_27}).  

Like in the classical case, see Eq.~(\ref{Introduction_35a}), in order to obtain the correlation function for the whole radiation  
field of a star, one has to sum over all individual pointlike regions over the entire surface of the star facing the observer:  
\ba
C_{\rm Star}\left(\vec{x}_A,\vec{x}_B\right) &=&
\frac{1}{A_{\rm star}^2}
\int_{A_{\rm star}} d^2 x_1\;
\int_{A_{\rm star}} d^2 x_2\;
\left| \Psi^{\rm Star}_{12} \right|^2 
\nonumber\\
\nonumber\\
&=& 1 + \frac{4\,L^2}{k^2\,d_{AB}^2\,R^2}\,J_1^2 \left(\frac{k\,R\,d_{AB}}{L}\right),
\label{Correlation_Function_12_E}
\ea
\noindent
which is obviously the same expression for the correlation function for stars in Eq.~(\ref{Introduction_35b}),  
previously obtained in the treatment in terms of classical electrodynamics.
As specific example, Eq.~(\ref{Correlation_Function_12_E}) reflects the analogon of Siegert relation 
in quantum electrodynamics, 
first obtained by Glauber, cf. Eq.~(10.26) in \cite{Glauber2}, and in general valid for incoherent light.

\subsubsection{Wavefunction for heavy-ion collisions}\label{WF_HIC}  

In case of heavy-ion collisions we have the following limits; see right pannel of Fig.~\ref{Wavefunction}:  
$L \gg d_{AB} \gg r_{12}$,  
where $L$ is the distance between fireball and detectors, $r_{12}$ is the distance among two pointlike regions on the  
fireball's surface and $d_{AB}$ is the distance between both detectors, see Fig.\ref{Electric_Fields}.  
In these limits, the photon wavefunction in (\ref{WF_5}) can considerably be simplified and is given by; cf.~\cite{Humanic}:  
\ba
\Psi_{12}^{\rm HIC} &=& \frac{1}{\sqrt{2}}
\bigg[
{\rm e}^{i\left[\mbox{\scriptsize\boldmath$k_A$}\cdot\left(\mbox{\scriptsize\boldmath$x_A$}-\mbox{\scriptsize\boldmath$x_1$}\right)
+ \mbox{\scriptsize\boldmath$k_B$}\cdot\left(\mbox{\scriptsize\boldmath $x_B$}-\mbox{\scriptsize\boldmath $x_2$}\right)\right]}
+ {\rm e}^{i\left[\mbox{\scriptsize\boldmath$k_A$}\cdot\left(\mbox{\scriptsize\boldmath$x_A$}-\mbox{\scriptsize\boldmath$x_2$}\right)
+ \mbox{\scriptsize\boldmath$k_B$}\cdot\left(\mbox{\scriptsize\boldmath $x_B$}-\mbox{\scriptsize\boldmath $x_1$}\right)\right]}
\bigg]\,,
\label{wave_function_HIC_5}
\ea

\noindent
where the wave-vector $\vec{k}_A$ is directed from the fireball's surface towards detector A,  
the wave-vector $\vec{k}_B$ is directed from the fireball's surface towards detector B.
The square of absolute value of this wavefunction is given by   
\ba
\left|\Psi_{12}^{\rm HIC}\right|^2 &=&
1 + \cos \left(k\,\frac{\vec{d}_{AB}\cdot\vec{r}_{12}}{L}\right), 
\label{wave_function_HIC_15}
\ea

\noindent
where $\vec{r}_{12} = \vec{x}_2 - \vec{x}_1$ is the vector from one surface-point $\vec{x}_1$ of the fireball of HIC to
another surface-point $\vec{x}_2$ of the fireball,    
$k= \left|\vec{k}_A\right| = \left|\vec{k}_A\right|$ is the absolute value of wave-vector of the photons, and 
we have used $\displaystyle \vec{k}_B-\vec{k}_A=k\,\frac{\vec{d}_{AB}}{L}$ which is valid for the geometry of heavy-ion collisions;  
see right pannel in Fig.~\ref{Wavefunction}. 

We note that (\ref{wave_function_HIC_15}) equals the corresponding classical expression obtained in (\ref{wave_function_HIC_14}).  
Like in the classical case, the absolute square of wavefunction in the limits of heavy-ion collisions  
in (\ref{wave_function_HIC_15}) agrees with the the absolute square of wavefunction in the limit for stars  
(\ref{wave_function_stars_15}). Let us recall here that the fireball created by a heavy-ion  
collision expands rapidly in time and space, hence the two-dimensional integration procedure in (\ref{Correlation_Function_12_E})  
has to be replaced by an invariant four-dimensional approach, which will briefly be discussed in the next Section.

\subsection{HBT effect for binary stars} 

\begin{figure}[!ht]
\begin{center}
\includegraphics[scale=0.3]{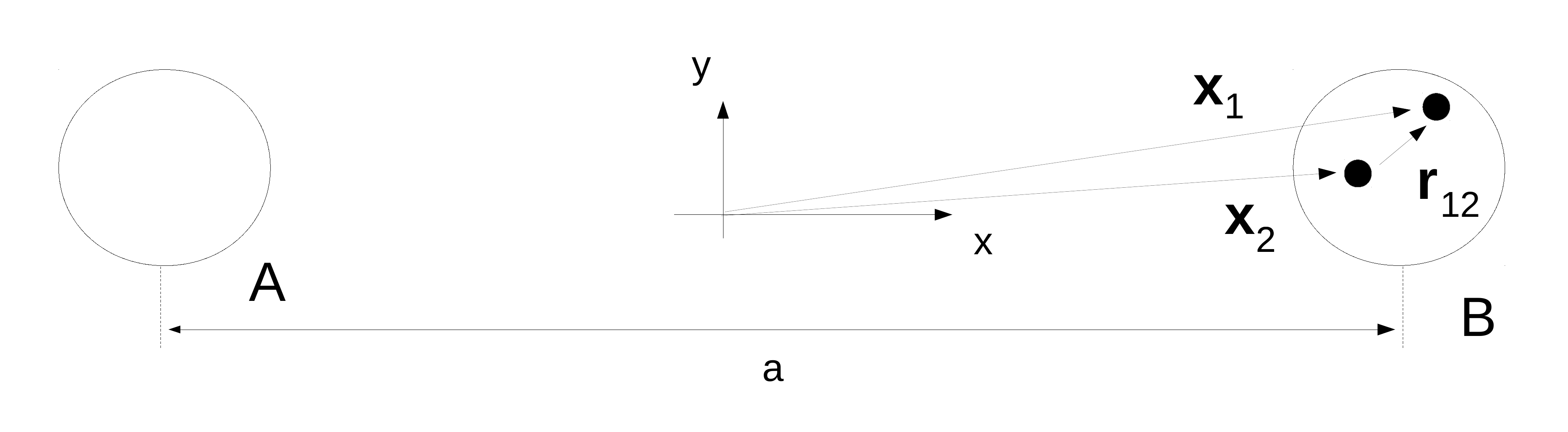}
\includegraphics[scale=0.3]{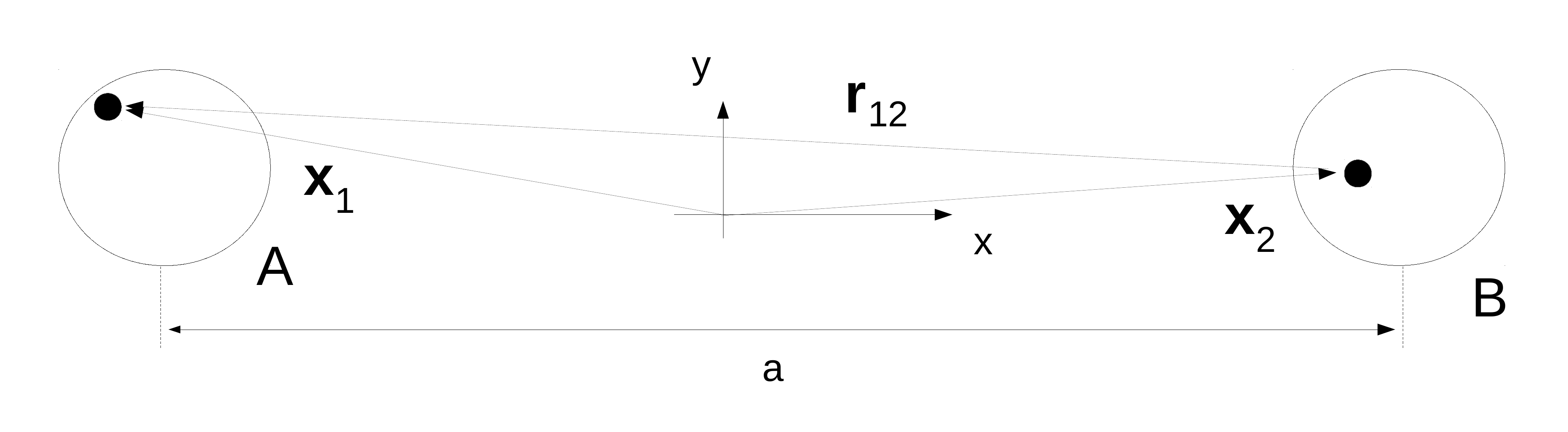}
\end{center}
\vskip -4mm
\caption{The both components of the binary system are denoted by $A$ and $B$.  
These figures elucidate the region of integration in 
Eq.~(\ref{Correlation_Binary_5}).}
\label{Integration_Region_1}
\end{figure}

In the previous sections it has been shown that classical 
electrodynamics and quantum-electrodynamical treatment  
leads to the same expressions for the correlation function 
of two pointlike sources, see Eq.~(\ref{Introduction_27}).  
Therefore, we can just apply 
any of these expressions in order to derive the correlation 
function for binary stars, given by 
\begin{eqnarray}  
C_{\rm Binary} \left(\vec{x}_A,\vec{x}_B\right) &=&
\frac{1}{2\,A_{\rm star}^2}\; \int d^2 x_1\; \int d^2 x_2\;  
\left[1 + \cos \left(k\,\frac{\vec{r}_{12} \cdot 
\vec{d}_{AB}}{L}\right)\right]\,, 
\label{Correlation_Binary_5}
\end{eqnarray}  

\noindent
where the regions of integrations are elucidated in
Fig.~\ref{Integration_Region_1}. 

\noindent
The integration in (\ref{Correlation_Binary_5}) yields for the 
correlation function for binary stars the following expression: 
\begin{eqnarray}
C_{\rm Binary} \left(\vec{x}_A,\vec{x}_B\right)
&=& 1 + \frac{4\,L^2}{k^2\,d_{AB}^2\,R^2}\,J_1^2 \left(\frac{k\,R\,d_{AB}}{L}\right)  
\left(1 + \cos \frac{k\,a\left(t\right)\,d_{AB}}{L}\right),  
\label{C_Binary_10}
\end{eqnarray}

\noindent
which is valid for $a\left(t\right) \ge 2\,R$; the extreme case $a\left(t\right) < 2\,R$ is not relevant for our investigation.
The term in (\ref{C_Binary_10}) which is not proportional to the cosine-function agrees with (\ref{Correlation_Function_12_E}),   
and describes the standard HBT effect for stars, that means photons from star A are correlated with each other 
and photons from star B are correlated with each other.  
The term in (\ref{C_Binary_10}) which is proportional to the  
cosine-function describes the HBT effect due to the binary system,  
that means photons from star A and photons from star B are correlated 
with each other. 

Two configurations are on the scope of our investigation. First, 
the case of inclination $i= 0$, where for circular orbits   
(eccentricity $e = 0$) the apparent distance between both stars 
as seen from the observer is simply given by  
\begin{eqnarray}
a\left(t\right) = A = {\rm const}\,.  
\label{C_Binary_14}
\end{eqnarray}

\noindent
Second, the case of edge-on binaries (inclination 
$\displaystyle i = \pi/4$), where for circular orbits the  
time-dependent distance $a\left(t\right)$ between both 
stars as seen from an observer takes the form   
\begin{eqnarray}
a\left(t\right) &=& A\,\cos \left(\frac{2\,\pi}{\mathcal{T}}\,\,t\right)\quad 
{\rm for} \quad a\left(t\right) \ge 2\,R. 
\label{C_Binary_15}
\end{eqnarray}

\noindent
Here, $t$ is the time, while $\mathcal{T}$ is the orbital period, given by
\begin{eqnarray}
\mathcal{T} &=& 2\,\pi\,\sqrt{\frac{A^3}{G\,\left(M_A + M_B\right)}}\,,  
\label{C_Binary_20}
\end{eqnarray}

\noindent
where $A$ is the semi-major axis,  
and $\displaystyle G= 6.67 \times 10^{-11}\,\frac{{\rm m}^3}{{\rm kg}\,
{\rm s}^2}$ is the gravitational constant.  

We will show that these results for the correlation function for 
stars (\ref{Correlation_Function_12_E}) and for binary stars 
(\ref{C_Binary_10}) can be reproduced by means of the more sophisticated 
approach of differential HBT-method,  
which has originally been introduced for analyzing the HBT effect in 
the fireball of heavy-ion collisions \cite{Differential_HBT_1,Differential_HBT_2,Differential_HBT_3}.  

\section{HBT in heavy-ion-physics and astrophysics}\label{Analogy}

By means of the HBT effect in astrophysics one may determine the spatial size of a star or the distance between 
the components of a binary system, while the HBT analysis in heavy-ion physics allows to get information about the space-time  
structure and evolution of the fireball created in heavy-ion collisions. But both approaches differ significantly.  
The HBT approach used in heavy-ion physics is more involved and more general than in case of astrophysics, simply because  
the hot and dense hadronic fireball expands rapidly in time and space on a timescale of 
about $10^{-22}\,...\,10^{-21}\,{\rm seconds}$,  
while a star emits continuously thermal radiation and the star's surface does not alter significantly over a very long period of  
time of about $10^{16}\,{\rm seconds}$. Therefore, a simple two-dimensional integration procedure like in (\ref{Introduction_35a}) 
is inapplicable in heavy-ion physics, instead a four-dimensional description is necessary. Both appraches are compared  
and the conditions are elucidated about how to modify the HBT analysis used in HIC for the case of astrophysics.  

\paragraph{HBT analysis in heavy-ion physics:}

Since the fireball expands rapidly, in heavy-ion physics the second-order correlation function in (\ref{Introduction_35a}) 
is generalized into a four-dimensional description in momentum-space, see e.g. \cite{WF10}, where it reads:  
\begin{eqnarray}
C\left(k_A, k_A\right) &=&
\frac{P_2\left(k_A, k_B\right)}
{P_1\left(k_A\right) P_1\left(k_B\right)}\;,
\label{Correlation_Function3}
\end{eqnarray}

\noindent
where $k_A = \left(\omega_{k_A}, \vec{k}_A\right)$ and $k_B = \left(\omega_{k_B}, \vec{k}_B\right)$ are the four-momenta 
of the two photons and $P_1$ and $P_2$ are the inclusive one-photon and  
two-photon distribution function, respectively. That means, $P_1\left(k_A\right)$ is the probability of detecting one photon of  
momentum $k_A$ in detector A or B, $P_1\left(k_B\right)$ is the probability of detecting one photon of momentum $k_B$ in detector 
A or B, and $P_2\left(k_A, k_B\right)$ is the probability of detecting two coincident photons of wavenumbers $k_A$ and $k_B$ in  
detectors A and B.
Let's assume the photons are emitted at two points on the fireball's surface, 
$x_1 = \left(t_1,\vec{x}_1\right)$ and $x_2 = \left(t_2,\vec{x}_2\right)$.  
Then, the one-photon and two-photon distribution function is then given by  
\begin{eqnarray}
P_1\left(k_A\right) &=& \int d^4 x_1 \; S\left(x_1, k_A\right), \quad  
P_1\left(k_B\right) = \int d^4 x_1 \; S\left(x_1, k_B\right),  
\label{Analogy_5}  
\\
\nonumber\\
P_2 \left(k_A, k_B \right) &=& \int d^4 x_1\, d^4 x_2 \; S\left(x_1, k_A\right) S\left(x_2, k_B\right) \left|\Psi_{12} \right|^2\,,  
\label{Analogy_10}
\end{eqnarray}

\noindent
where $\Psi_{12}$ is the two-photon wavefunction in (\ref{WF_5}) and its absolute square in (\ref{WF_1}),  
and the functions $S\left(x_1, k_A\right)$ and $S\left(x_2, k_B\right)$ are the socalled photon emission functions,  
a term originally introduced in heavy-ion physics and which will be considered in the next Section.  

In heavy-ion collisions the HBT effect was originally been used for the same purpose as in astrophysics,   
namely the determination of the system size by measuring the correlations of bosons (pion pair correlations and later 
photon pair correlations) which are emitted from the expanding fireball created by the heavy-ion collision \cite{FirstHBTs}.  
Up to now, the HBT analysis remains the only experimental and model-independent approach to determine the size af 
the expanding fireball. Ever since, the effect of photon (or pion correlations) has become a fundamental aspect  
not only in field of quantum optics, but also in the theory and experiment of heavy-ion (or nucleon-nucleon) collisions.  
That means, similar second-order correlation functions (\ref{Correlation_Function3}) have also been measured in  
heavy-ion collisions for pion-correlations \cite{Experiment1,Experiment2} (where $\Psi_{12}$ would then be the two-pion  
wavefunction and the four-dimensional integrations would run over the fireball).  

Todays highly sophisticated high energy heavy-ion experiments can even detect the ellipsoidal shape of the fireball  
and also its tilt by means of the HBT effect \cite{MLisa,DM95}. At present collision energies the  
angular momentum and rotation of the hot and dense fireball is becoming a
dominant feature of heavy-ion collisional processes, and the possibility to
analyze such effects by the so-called differential HBT method was subject of
the investigations in \cite{Differential_HBT_1,Differential_HBT_2,Differential_HBT_3}.  

In heavy-ion reactions numerous particles are registered, of the order of thousand, in every single collision event. The  
timespan of the reaction is of the order of $10\,{\rm fm}/c$ just as the size of the system. So, the particles emitted and  
observed in one collision event are considered as contemporary, and interacting with each other.  Consequently an event by event
measurement registers an M-particle correlation (where M is the bserved multiplicity of the event). For the second-order correlation  
function every pair is selected from the M particles and this provides the probability distribution $P_2\left(k_A, k_B\right)$ 
which allows to determine the second-order correlation function in (\ref{Correlation_Function3}).  

\paragraph{HIB analysis in astrophysics:} 

In order to exploit the advanced machinery of HBT analysis developed in heavy-ion physics, in our subsequent investigation we will
start to consider the correlator in (\ref{Correlation_Function3}) instead (\ref{Introduction_35a}).
The results of such an approach will finally yield similar expressions for the correlator as given by 
Eq.~(\ref{Correlation_Function_12_E}) for stars and by Eq.~(\ref{C_Binary_10}) for binaries.  
In doing so, several aspects have carefully to be accounted for.  

First, as mentioned above, in case of stars and binary stars we face a continuous emission of thermal radiation in time.  
We have to assume that the luminosity from the stars is big enough, so that in a correlation time period we have sufficient  
amount of photons emitted in the direction of the observer so that the observable multiplicity, $M \gg 4$. According to  
the investigations in \cite{Photon_Flux} we can savely assume that this requirement is obeyed for many stars located  
in the neighborhood of the Solar system; e.g. from Sirius $10^6 {\rm photons}/{\rm cm}^2 \,{\rm sec}$ arrive the Earth's surface. 
Then, the subsequent analysis can be repeated in every such time interval, similarly to subsequent collision events in heavy-ion  
reactions. However, in case of faint and optically not resolvable binaries, the luminosity might not be too large, so it is also  
important that the detection efficiency should be as high as possible. In this respect we mention that most modern optical 
technologies allow to generate CCD detectors with a plane of about $1\,{\rm square-meter}$, as used for instance in the ESA 
astrometry mission Gaia \cite{Gaia}.  

Second, we have to account for the fact, that the opacity of a star is so extremely high that the visible photons do not  
originate from inner regions of the star but from the thin photosphere of the star, that is to say from the two-dimensional surface.  
Accordingly, the HBT approach used in heavy-ion physics has to be modified for the case of astrophysics,  
by simplifying the four-dimensional integrals in (\ref{Analogy_5}) - (\ref{Analogy_10}) by integrals which run over
the two-dimensional hyper-surface of stars. 

Third, the wavefunction in (\ref{WF_5}) is valid both for the geometry  
of HIC (i.e. $d_{AB} \gg R$ and $R \ll L$) and for the geometry of astrophysics (i.e. $d_{AB} \ll R$ and $R \ll L$),  
and will be simplified and designed for the specific case of stars, cf. considerations in Section \ref{HBT}.  
These issues will be the subject of Section \ref{TEF} and Section \ref{PCF}. 

\section{The Emission Function}\label{TEF}

Let us consider the photon emission function $S\left(x, k\right)$ which is a fundamental ingredient in the 
correlation functions in (\ref{Correlation_Function3}) via Eqs.~(\ref{Analogy_5}) - (\ref{Analogy_10}).  
The photon emission function, frequently called sourcefunction, is the probability distribution of emitting a photon from the  
source point $x^{\mu}=\left(ct,\vec{x}\right)$ with four-momenta $k^{\mu} = \left(k^0,\vec{k}\right)$, that means the number of  
photons, $\Delta N$, emitted in the phase-space element $\Delta^3 x\,\Delta^3 k$ per unit time $\Delta t$.  
A Lorentz invariant scalar can be obtained by multiplying the photon-energy $k^0 = \omega_k$ of the emitted photons:  
\begin{eqnarray}
S\left(x,k\right) &=& k^0 \frac{\Delta N}{\Delta t \,\Delta^3 x\,\Delta^3 k}\,.
\label{Source_Funtion1}
\end{eqnarray}

\noindent
Accordingly, the total number of photons is given by the summation over the entire space-time and momentum-space: 
\begin{eqnarray}
N &=& \int d^4 x \int \frac{d^3 k}{k^0} S\left(x,k\right). 
\label{Source_Funtion2}
\end{eqnarray}
 
\noindent
For the calculation of the photon emission function in (\ref{Source_Funtion1}) one needs the photon four-current, 
defined by \cite{Cs-0}:  
\begin{eqnarray} 
N^\mu(x) &=& \int k^\mu \frac{d^3 k}{k^0} f(x,k)\,,
\end{eqnarray} 

\noindent
where $f(x,k)$ is the invariant scalar phase-space density distribution (black body distribution) of the emitted photons,  
\begin{eqnarray}
f\left(x,k\right) &=&  \frac{g_{\gamma}}{\left(2\pi\right)^3}
\frac{1}{\exp\left(k^\mu\,u_\mu/T\right)-1}\,,
\label{black_body_distribution}
\end{eqnarray}

\noindent
where the degeneracy-factor $g_{\gamma}=2$ accounts for the two polarizations of the photons.
This single particle distribution depends on the local four-velocity $u^\mu(x)$ of the light-source.
In order to obtain the total number of photons crossing a space-time hyper-surface $\Sigma$, 
we have to integrate over all momenta and
over the entire surface:
\begin{eqnarray}
N &=& \int_{\Sigma}\left[\int\frac{d^3 k}{k^0}\,k^{\mu}\,
f\left(x,k\right) \right] d \Sigma_{\mu}\,,
\label{Photon_Number}
\end{eqnarray}

\noindent
where $d\Sigma_{\mu}$ is an infinitesimal three-dimensional surface-element. 
By not performing the momentum integral in  
(\ref{Photon_Number}) we recover the invariant triple differential cross section by the Cooper-Frye formula \cite{CF}:  
\begin{eqnarray}
E \frac{dN}{d^3 k} &=& \int_\Sigma f\left(x,k\right)\ k^\mu d\Sigma_{\mu}\,,
\label{CF}
\end{eqnarray}

\noindent
with $E=k^0$. On account of Eqs.~(\ref{Source_Funtion2}) and (\ref{Photon_Number}), we get an integral relation between 
photon sourcefunction $S\left(x,k\right)$ and photon density distribution $f\left(x,k\right)$ as follows:  
\begin{eqnarray}
\int d^4 x \, S\left(x,k\right) &=& \int_{\Sigma}\,f\left(x,k\right)\,k^{\mu}\,d \Sigma_{\mu}\,.   
\label{Relation_S_f}
\end{eqnarray}

\noindent
In the r.h.s. of Eqs.~(\ref{Photon_Number}) - (\ref{Relation_S_f}) it is assumed that the emission takes place through a 
three-dimensional hyper-surface $\Sigma$ with the normal vector $d\Sigma_\mu = d\Sigma_\mu(x)$. 
The hyper-surface is bordered by the outgoing light-cone, so all emitted particles should cross the hyper-surface.  
We are interested in the hypersurface $x^0={\rm const.}$. Then, for time-like surface we have 
$d\Sigma_\mu = dx\,dy\,dz\,\hat{\sigma}_\mu$, with $\hat{\sigma}_\mu$ being a dimensionless unit four-vector.  
We will assume the simplest case for the time-like normal $\hat{\sigma}_\mu=\left(1,0,0,0\right)$, but actually 
we do not need this requirement  
to determine correlation function since the invariant scalar $k^{\mu}\,\hat{\sigma}_{\mu}$ cancels in the normalized correlator.  
Then, according to relation (\ref{Relation_S_f}) and in virtue of Eq.~(\ref{black_body_distribution}) 
the source-function can be parametrized as follows; cf.~\cite{Differential_HBT_1,Source_Function}:
\begin{eqnarray}
S\left(x,k\right) &=& \frac{2}{\left(2\pi\right)^3}\,
\frac{\gamma_s}{\exp\left(k^\mu\,u_\mu\left(x\right)/T\left(x\right)\right)-1}\,H\left(t\right)\,
G\left(x,y,z\right)\,k^\mu\,\hat{\sigma}_\mu(x)\,,  
\label{Source_Function_A}
\end{eqnarray}

\noindent
where $\gamma_s = \sqrt{1-v_s^2}$ is the Lorentz factor with $v_s = \left|\vec{v}_s\right|$ being the absolute value of 
spatial velocity of the source; let us recall that the chemical potential for photons is zero (i.e. fugacity factor equals $1$).  
The function $H\left(t\right)$ governs the time-dependence of photon emission, which in case of 
heavy-ion collisions is usually described by a delta-function (sudden freeze-out) or an exponentially decreasing 
Gauss-function (gradually freeze-out), e.g. \cite{Differential_HBT_1}, because 
the fireball emits photons during a small time-intervall around the freeze-out time $t_0$.  
Here, in astrophysical systems, the thermal light-source (star, binary star) permanently emits photons, that means  
the number $N$ of emitted photons is proportional to time $t$, i.e.: $N \sim t$), hence the function $H\left(t\right)$  
in (\ref{Source_Function_A}) remains basically a constant over a very long period of time, thus differs considerably 
from the case of HIC scenarios.  
The function $G\left(x,y,z\right)$ is the space-time emission density across the layer of the hypersurface.  
For stars the layer is narrow, i.e. the diameter of the photosphere of a solar-like star is much smaller 
than the diameter of the star, hence the function $G\left(x,y,z\right)$   
can be simplified by a two-dimensional surface and diameter of the layer in z-direction is described by a delta-function. 
Using a Gaussian profile for a star with mean-radius $R$, we have  
\begin{eqnarray}
G\left(x,y,z\right) &=& \delta\left(z\right)\,{\rm exp} \left( - \frac{x^2 + y^2}{2\,R^2}\right),  
\label{Gaussian_Profile}
\end{eqnarray}

\noindent
and the Boltzmann approximation (J{\"u}ttner-distribution), 
we finally arrive at the sourcefunction for a two-dimensional surface of a star with radius $R$:  
\begin{eqnarray}
S\left(x,k\right) &=& \frac{\gamma_s}{C_{\gamma}}\,\exp\left(- \frac{k^\mu\,u_\mu\left(x\right)}{T\left(x\right)}\right)\,
H\left(t\right)\,\delta\left(z\right)\,{\rm exp}\left(- \frac{x^2 + y^2}{2\,R^2}\right)\,
k^\mu\,\hat\sigma_\mu(x)\,, 
\label{Source_Function_D}
\end{eqnarray}
 
\noindent
where $C_{\gamma}=4\,\pi^3$. In what follows we will use this expression for the photon emission function.
The generalization for the case of two light-sources is straightforward.

\section{ The Correlation Function}\label{PCF}

The correlation function in Eq.~(\ref{Correlation_Function3}) is defined as the inclusive two-photon distribution, normalized  
by dividing with by the product of the inclusive one-photon distributions. The correlation function in momentum-space depends 
on $k_A$ and $k_B$ being the photon four-momenta detected at detector A and B, respectively. We will assume photons are emitted 
at two points, $\vec{x}_1$ and $\vec{x}_2$ on the stars's surface, which are separated in space, but due to the distant source 
we cannot resolve where the photons are coming from. 

\subsection{The wavefunction for stars}\label{PCF_1}

In order to determine the two-photon distribution in (\ref{Analogy_10})   
one needs to implement the two-photon wavefunction in (\ref{WF_5}), which can be written as follows 
\footnote{where ($k_1=k_2$): $\displaystyle \vec{k}_{1A} = k\,\frac{\vec{x}_A - \vec{x}_1}{L}$, 
$\displaystyle \vec{k}_{1B} = k\,\frac{\vec{x}_B - \vec{x}_1}{L}$,
$\displaystyle \vec{k}_{2A} = k\,\frac{\vec{x}_A - \vec{x}_2}{L}$,
$\displaystyle \vec{k}_{2B} = k\,\frac{\vec{x}_B - \vec{x}_2}{L}$.
}: 
\ba
\Psi_{12} &=& \frac{1}{\sqrt{2}}
\bigg[
{\rm e}^{i\left[\mbox{\scriptsize\boldmath$k_{1A}$}\cdot\left(\mbox{\scriptsize\boldmath$x_A$}-\mbox{\scriptsize\boldmath$x_1$}\right)
+\mbox{\scriptsize\boldmath$k_{2B}$}\cdot\left(\mbox{\scriptsize\boldmath $x_B$}-\mbox{\scriptsize\boldmath $x_2$}\right)\right]}
+{\rm e}^{i\left[\mbox{\scriptsize\boldmath$k_{1B}$}\cdot\left(\mbox{\scriptsize\boldmath$x_B$}-\mbox{\scriptsize\boldmath$x_1$}\right)
+\mbox{\scriptsize\boldmath$k_{2A}$}\cdot\left(\mbox{\scriptsize\boldmath $x_A$}-\mbox{\scriptsize\boldmath $x_2$}\right)\right]}
\bigg]\,,
\label{WFss}
\ea

\noindent
Recall, the Boson wavefunction in (\ref{WFss}) is fully symmetric, irrespectively if we symmetrize for 
$\vec{x}_A \leftrightarrow \vec{x}_B$ or $\vec{x}_1 \leftrightarrow \vec{x}_2$, and is valid for heavy-ion collisions and stars.  
We can consider that $\vec{k} = \frac{1}{4}(\vec{k}_{1A} + \vec{k}_{2B} + \vec{k}_{1B} + \vec{k}_{2A})$.  
Then the three-momentum vectors become (note that these vectors are not independent of each other: 
$\vec{k}_{1A} - \vec{k}_{1B} = \vec{k}_{2A} - \vec{k}_{2B}$),  
\ba 
\vec{k}_{1A} &=& \vec{k} - \frac{\vec{\kappa}}{2} + \frac{\vec{q}}{2}\,, \quad  
\vec{k}_{2B}  = \vec{k} + \frac{\vec{\kappa}}{2} - \frac{\vec{q}}{2}\,, 
\label{k-s_1}
\\ 
\nonumber\\ 
\vec{k}_{1B} &=& \vec{k} - \frac{\vec{\kappa}}{2} - \frac{\vec{q}}{2}\,,\quad 
\vec k_{2A}  =  \vec{k} + \frac{\vec{\kappa}}{2} + \frac{\vec{q}}{2}\,, 
\label{k-s_2}
\ea

\noindent
where 
\ba
\vec{\kappa} &=& k\,\frac{\vec{x}_1 - \vec{x}_2}{L} \equiv \frac{\vec{r}_{12}}{L}  
\quad {\rm and} \quad \vec{q} = k\,\frac{\vec{x}_A - \vec{x}_B}{L} \equiv k\,\frac{\vec{d}_{AB}}{L}\,.  
\ea

\noindent
In case of stars we have $d_{AB} \ll r_{12}$, and the four $\vec k$-vectors can be expressed as:
\ba
\vec{k}_{1A} &=& \vec{k} - \frac{\vec{\kappa}}{2}\,, \quad \vec{k}_{2B} = \vec{k} + \frac{\vec{\kappa}}{2} \,, 
\label{k-ss_1}
\\
\nonumber\\
\vec{k}_{1B} &=& \vec{k} - \frac{\vec{\kappa}}{2} \,, \quad \vec{k}_{2A} = \vec{k} + \frac{\vec{\kappa}}{2} \,.   
\label{k-ss_2}
\ea

\noindent
With these parameters the wavefunction becomes
\begin{eqnarray} 
\Psi_{12}^{\rm Star} &=& \frac{1}{\sqrt{2}}
\exp\left[(i \vec{k} \cdot (\vec{x}_A + \vec{x}_B - \vec{x}_1 - \vec{x}_2) \right] 
\nonumber\\ 
&& \hspace{-1.5cm} \times \left(
\exp\left[- i\,\frac{\vec \kappa}{2} \cdot  
(\vec x_A{-}\vec x_B{-}\vec x_1{+}\vec x_2) \right] 
+ \exp\left[ + i\,\frac{\vec \kappa}{2} \cdot 
(\vec x_1{-}\vec x_2{+}\vec x_A{-}\vec x_B ) \right] 
\right).
\label{W-1}
\end{eqnarray} 

\noindent
Now in order to be able to perform the integrals over the source variables, $\vec x_1$ and $\vec x_2$, we insert the variable  
$\vec{\kappa} = k (\vec{x}_1 - \vec{x}_2)/L$  and we describe the remaining observable parameters in terms of $\vec q$ as  
$\vec d_{AB} = L\,\vec{q}/k$, and then we get  
\begin{eqnarray} 
\Psi_{12}^{\rm Star} &=& \frac{1}{\sqrt{2}}
\exp\left[i \vec{k} \cdot (\vec x_A + \vec x_B - \vec x_1 - \vec x_2) \right]  
\nonumber\\ 
&& \hspace{-2.0cm} \times \left(
\exp\left[- \frac{i\,k}{2L}(\vec x_1 - \vec x_2) \cdot 
\left(\frac{L\vec q}{k}{-}(\vec x_1 - \vec x_2)\right) \right]  
+ \exp\left[+\frac{i\,k}{2L}(\vec x_1 - \vec x_2) \cdot  
\left(\frac{L\vec q}{k}{+}(\vec x_1 - \vec x_2)\right) \right] 
\right) \,,
\nonumber\\ 
\label{W-2}
\end{eqnarray} 

\noindent
which, after performing the multiplications in the exponents of the last two terms, becomes  
\begin{eqnarray} 
\Psi_{12}^{\rm Star} &=& \frac{1}{\sqrt{2}}
\exp\left[i \left(\vec k \cdot (\vec x_A + \vec x_B - \vec x_1 - \vec x_2) +
\frac{k}{L}(\vec x_1 - \vec x_2)^2\right)\right]  
\nonumber\\  
&& \times \Bigg[
\exp\left[+ i \left(\frac{\vec q}{2} \cdot (\vec x_1{-}\vec x_2)\right)\right] +
\exp\left[- i\left(\frac{\vec q}{2} \cdot (\vec x_1{-}\vec x_2)\right)\right] 
\Bigg].
\label{W-3ss}
\end{eqnarray} 

\noindent
The structure of this function is the same as of Eq.~(\ref{WF_5}), and
then for the absolute value square of the wavefunction we obtain: 
\ba 
\left|\Psi^{\rm Star}_{12}\right|^2 &=& 1 + \frac{1}{2} \bigg( {\rm exp}\left[+ i \vec q \cdot (\vec x_2-\vec x_1)\right]  
+ {\rm exp}\left[- i \vec q \cdot (\vec x_2-\vec x_1) \right] \bigg)  
\nonumber\\
\nonumber\\
&=& 1 +  \cos \left( k \frac{\vec{d}_{AB} \cdot \vec{r}_{12}}{L} \right)\,, 
\label{W-4ss}
\ea 

\noindent
where we have used the relation $ \vec{q} = k\,\vec{d}_{AB}/L$. The result in (\ref{W-4ss})  
agrees with the earlier obtained square of the wavefunction in Section \ref{QED}, cf. Eq.~(\ref{wave_function_stars_15}).  
Thus we express $\vec{\kappa}$ in terms of the initial state positions $\vec \kappa = k (\vec x_1 - \vec x_2)/L$,  
and we perform the integral over the source points included in $\vec{\kappa}$ also.  

\subsection{The correlation function}\label{PCF_2}

With the result for the absolute value of wavefunction in (\ref{W-4ss}), we obtain for the one-photon distribution:  
\begin{eqnarray}  
P_1(k) &=&  \int d^4 x\ S(x, k),  
\label{P1b}
\end{eqnarray}  

\noindent
while the two-photon distribution becomes:  
\begin{eqnarray}  
P_2\left(k+\frac{\kappa}{2} + \frac{q}{2} , k - \frac{\kappa}{2} - \frac{q}{2}\right) &=& \int d^4 x_1\, d^4 x_2\,
S\left(x_1,k+\frac{k x_1}{2\,L}+\frac{q}{2}\right)\,S\left(x_2,k-\frac{k\,x_2}{2\,L}-\frac{q}{2}\right)  
\nonumber\\ 
&& \hspace{-1.0cm} \times \left[1 + \frac{1}{2} \bigg( {\rm exp}\left[i \vec q \cdot (\vec x_2-\vec x_1)\right]
+ {\rm exp}\left[- i \vec q \cdot (\vec x_2-\vec x_1) \right] \bigg) \right]. 
\label{P2c}
\end{eqnarray}  

\noindent
Using Eqs.~(\ref{P1b}) and (\ref{P2c}) with the wavefunction in Eq.~(\ref{W-4ss}),  
together with the definition of the correlation function, we have:
\begin{eqnarray}
C(k,q)&=& 1 + \frac{R(k,q)}{\left| \int d^4 x\,  S(x, k) \right|^2}\,,  
\label{Cmain-ss}
\end{eqnarray}
where
\ba 
R(k,q) &=& {\rm Re}\, \left[J_{(+)}(k,q)\ J_{(-)}(k,-q) \right].  
\label{R-defss}
\ea 

\noindent
Here $R(k,q)$ can be calculated via the function
\ba
J_{(\pm)}(k,q) &=& \int d^4x\ S(x,k(1{\pm}x/2L))\, \exp(iqx)  
\nonumber\\
&=& \int d^4x\ S(x,k(1{\pm}x/2L)\, [\cos(qx) + i \sin(qx)] \,,
\label{J-defss}
\ea
and we obtain the $R(k,q)$ function as

\ba 
R(k,q) &=&  \int d^4 x_1\,d^4 x_2\,\cos\bigg(\vec{q}\cdot\left(\vec{x}_1-\vec{x}_2\right)\bigg)\,
S\left(x_1, k+\frac{k\,x_1}{2\,L}\right)\,S\left(x_2, k-\frac{k\,x_2}{2\,L}\right).  
\label{R1ss}
\ea 

\noindent
This can easily be verified, by using Eq.~(\ref{J-defss}), forming a double integral over $d^4x_1\, d^4x_2$ from  
$J_{(+)}(k,q)\, J_{(-)}(k,-q)$, yielding to a term $\exp[-iq(x_1-x_2)]$. Then  
taking the real part of the double integral leads to a term $\cos[q(x_1-x_2)]$ and this recovers Eq.~(\ref{P2c}).

\subsection{Source with Black Body J{\"u}ttner-distribution}\label{PCF_3}

Let us consider the $S(x_1, k_1)\, S(x_2, k_2)$ term in Eq.~(\ref{P2c}). According to Eq.~(\ref{Source_Function_D}), we assume  
that the single photon distributions, $f(x,k)$, in the source function are J{\"u}ttner distributions, which depend on the  
local velocity, $u^\mu(x)$, via the term:   
\ba 
&~& \exp\left[\frac{- k^\mu\,u_\mu(x)}{T(x)}\right]\,. 
\ea 

\noindent
For a normal star the surface temperature is of order $6000K$, and for optical light in nano meter range the exponential 
is of order $10^2$, so that the J{\"u}ttner distribution is a proper approximation. 
In general, the local flow velocity might be different in different locations, $x_1$ and $x_2$, and 
this fact influences the correlations of the observed momenta. Thus, the scalar products in terms of $k$ and $\kappa$ are: 
\ba 
{\rm e}^{-k_1\,u_1}\,{\rm e}^{-k_2\,u_2}  
&=& {\rm e}^{-(k+\kappa/2)\,u_1}\,{\rm e}^{-(k-\kappa/2)\,u_2}  
= {\rm e}^{-k\,u_1}\,
{\rm e}^{-k\,u_2}\,{\rm e}^{-\kappa\,u_1/2}\,{\rm e}^{+\kappa u_2/2}\,,  
\label{k-kappa}
\ea 

\noindent
where we used the notation $u_1 = u(x_1) = u^\mu(x_1)$. We assume that for a given detector position the normal direction of the  
emission is approximately the same, so for the two sources the term $k^\mu  \hat\sigma_\mu(x)\,$ is the same and it  
cancels in the nominator and denominator. Thus, the expression of the $J$-function in Eq.~(\ref{J-defss}) will be modified to  
\be
J_{(\pm)}(\vec k,\vec q) =  \int d^4x\ S(\vec x,\vec k)\, 
\exp\left[ \pm\, \frac{k\, \vec x \cdot \vec u(x)}{2\,L\,T(x)} \right]\, 
\exp(i\vec q \cdot \vec x)\ ,
\label{J2ss} 
\ee

\noindent
and subsequently we can calculate $R$ in Eq.~(\ref{R-defss}),
\be
R(k,q)=\int d^4 x_1\,d^4 x_2\, S(x_1,k)S(x_2,k)
\exp\left[- \left(\frac{\kappa_1\,u_1}{2\,T(x_1)}-\frac{\kappa_2\,u_2}{2\,T(x_2)}\right)\right]  
\cos[\vec{q}\cdot\left(\vec{x}_1-\vec{x}_2\right)]\,, 
\label{R-u}
\ee

\noindent
where $\kappa_1 = \kappa(x_1), \kappa_2 = \kappa(x_2)$,  
$ \vec \kappa(\vec x) = k \vec x/(2L)$, and with the correlation function in Eq.~(\ref{Cmain-ss}).  
The Eqs.~(\ref{J2ss}) and (\ref{R-u}) are consistent with the definition 
(\ref{R-defss}) of $R$ in terms of $J_{(\pm)}$. One has to keep in mind 
that the integrals in Eq.~(\ref{R-u}) are extended over 
$\vec \kappa(\vec x_1)$ and $\vec \kappa(\vec x_1)$ also.
If we have few point like sources the integral becomes a sum over the
sources, and then the $\kappa_i$ values should be taken at the
same position as the arguments of the velocities, $u(x_i)$.

\section{One Source}\label{SSofcs}

\subsection{One source at rest}

We will determine the correlation function for one source as given by Eq.~(\ref{Cmain-ss}). 
First we consider the invariant scalar $k^\mu u_\mu$, which can be calculated in the frame
where the surface-element is at rest. We have then $u^\mu = (1,0,0,0)$ for the four-velocity, hence  
\begin{eqnarray}
k_\mu\,u^\mu &=& k^0\,,  
\label{Four_Velocity_One_Source} 
\end{eqnarray}

\noindent
with $k^0=E_k$ being the energy of one photon-mode in the rest frame of the star. 
According to (\ref{Source_Function_D}) and (\ref{Four_Velocity_One_Source}), the source function for one source at rest reads 
\begin{eqnarray}
S\left(x,k\right) &=& \frac{1}{C_{\gamma}}\,\exp\left(- \frac{E_k}{T_s}\right)\,
H\left(t\right)\,\delta\left(z\right)\,{\rm exp}\left(- \frac{x^2 + y^2}{2\,R^2}\right)\,
k^\mu\,\hat\sigma_\mu(x)\,,
\label{Source_Function_one_source_1}
\end{eqnarray}

\noindent
where $T_s$ is the temperature of the source.   
The denominator in (\ref{Cmain-ss}) is the single photon distribution for which we obtain:
\ba
\int d^4x\, S(x, k) &=& \frac{D}{C_{\gamma}}\,(k^\mu \hat\sigma_\mu)
\exp\left(-\frac{E_k}{T_s}\right)
\int_{-\infty}^{+\infty} e^{-\frac{x^2}{2R^2} } dx
\int_{-\infty}^{+\infty} e^{-\frac{y^2}{2R^2} } dy
\nonumber\\
&=& 2\,\pi\,R^2\,\frac{D}{C_{\gamma}}\,(k^\mu \hat\sigma_\mu)\,\exp\left(-\frac{E_k}{T_s}\right)\,, 
\label{InS}
\ea

\noindent
where we have defined $D= \int dt\,H\left(t\right)$ (in the normalized correlator (\ref{Cmain-ss}) this factor cancels out), 
and we have used $\int dz\,\delta\left(z\right)=1$ as well as:    
\begin{eqnarray}
\int_{-\infty}^{+\infty} e^{-a x^2} d x &=& \sqrt{\frac{\pi}{a}}\,.  
\end{eqnarray}

\noindent
In this simplest case we also assume that the surface direction is
$\hat\sigma^\mu = (1, 0, 0, 0)$ where the
observer is located in the z-direction.
The nominator in (\ref{Cmain-ss}) is determined by means of Eq.~(\ref{J2ss}), and we obtain 
\ba
J_{(\pm)}(k,q) &=& \int d^4x\, e^{iqx} \,e^{\mp{\kappa}^0/(2T_s)} S({x}, k)
\nonumber\\
&=& \frac{D}{C_{\gamma}}\,(k^\mu \hat\sigma_\mu)\, 
e^{-\frac{E_k \pm \kappa^0/2}{T_s}}
\int_{-\infty}^{+\infty}\,e^{{-\frac{x^2}{2R^2}}} e^{-iq_x x} dx
\int_{-\infty}^{+\infty}\, e^{{-\frac{y^2}{2R^2}}}\, 
e^{-i q_y y} dy 
\nonumber\\
&=& 2\,\pi\,R^2\,\frac{D}{C_{\gamma}}\,(k^\mu \hat\sigma_\mu)\,  
\exp\left[{-}\frac{E_k}{T_s}\right]
\exp\left[{\mp}\frac{{\kappa}^0}{2T_s}\right]
\exp\left[-\frac{R^2}{2} q^2\right] \,,
\ea

\noindent
where we used
$\int_{-\infty}^\infty \exp(-p^2x^2 {\pm} qx) dx =
(\sqrt{\pi} / p) \times \exp(q^2/(4p^2)) $ \cite{GR}.

In the $J(k,q)_{(+)} J(k,-q)_{(-)}$ product the 
terms $\exp[\pm {\kappa}^0 /(2T_s)]$
cancel each other. 
Inserting these equations into (\ref{Cmain-ss}) we get for the correlation function for one source  
\begin{eqnarray}
C\left(k, q\right) &=& 1 + \exp\left(- R^2 q^2\right).  
\label{Csss}
\end{eqnarray}

\noindent
The correlation function for one source at rest does not dependent on $k$. 
One might want to compare the correlation function in (\ref{Csss}) with Eq.~(\ref{Introduction_35b}). 
However, one should not wonder about the slight difference, because in order to obtain (\ref{Csss}) we have 
used a Gaussian source function in (\ref{Gaussian_Profile}), while in (\ref{Introduction_35b})  
the Einstein-Hopf model for the thermal radiation source has been applied.

\begin{figure}[!ht]
\begin{center}
\includegraphics[scale=0.4]{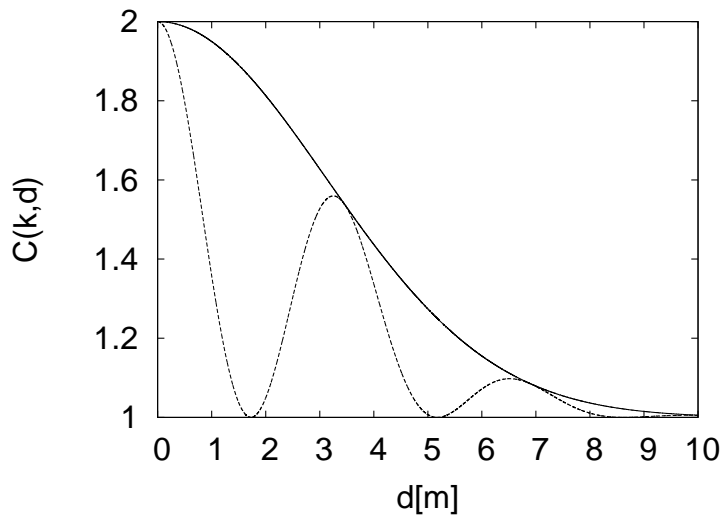}
\end{center}
\vskip -4mm
\caption{The correlation function for one source in Eq.~(\ref{Csss}) (solid line) and
the correlation function or two spherical Gaussian sources in Eq.~(\ref{C2}) (dotted line).  
The parameter are: $\textit{R} = 1.2\cdot10^9$m,
$\textit{k} =7.85\cdot10^6$m$^{-1}$,
$\textit{L}=4.14\cdot10^{16}{\rm m}$,
$x_s=4R$.} 
\label{One_Source}
\end{figure}

As Fig.~\ref{One_Source} shows, the correlation for such a star size yields an extended  
distribution in terms of the detector distance $d_{AB}$, which 
corresponds to $q_{AB}$ as $ q_{AB} = k\,d_{AB}/L$. Because of the symmetry of the source 
the correlation function only depends on the absolute value of $q_{AB}$, so 
there is no preferred direction in the plane perpendicular to the source. 
The dependence of the source distance to the correlation function through $q_{AB}$ 
limits the region where the correlation function will be applicable. 
Systems under investigation are bound to the near-zone of the Solar system.

\subsection{One source in motion}

Let us now consider one source which moves
in the z-direction with a velocity $v_z$. Then we have,
$  u_s^\mu = \gamma_s (1,0,0,v_z) $, and the scalar product
$k\cdot u_s/T_s = k_\mu u_s^\mu / T_s $ provides an additional
contribution to the correlation function. However, in the case
of a single star the
velocity and the temperature do not change within
the star, so the modifying term in
Eq.~(\ref{R-u}) becomes unity, and we have  
\begin{eqnarray}
k_\mu\,u^\mu &=& \gamma_s \left(E_k - k_z\,v_z\right). 
\label{Four_Velocity_One_Source_moving} 
\end{eqnarray}

\noindent
According to (\ref{Source_Function_D}) and (\ref{Four_Velocity_One_Source_moving}), the source function for one moving source 
\begin{eqnarray}
S\left(x,k\right) &=& \frac{\gamma_s}{C_{\gamma}}\,H\left(t\right)\,\delta\left(z\right)\,
\exp\left(- \frac{\gamma_s\,\left(E_k - k_z\,v_z\right)}{T_s}\right)\,
{\rm exp} \left(- \frac{x^2 + y^2}{2\,R^2}\right)\,
k^\mu\,\hat\sigma_\mu(x)\,. 
\label{Source_Function_one_source_2}
\end{eqnarray}

\noindent
Within the source the velocity $u_s$ and temperature
$T_s$ are assumed to be the same.  The spatial 
integrals can be performed in the
rest frame of the source, giving the same integral result as above 
(\ref{InS}), because the moving cell-size shrinks, but the apparent
density increases, so that the total number of photons in a 
cell remains the same as it is an invariant scalar. 
For the integral of the one-photon contribution we obtain  
\ba
\int d^4 x\,  S({x}, k) &=& 2\,\pi\,R^2\,\gamma_s \,\left(k^\mu \hat\sigma_\mu\right) \,\frac{D}{C_\gamma} 
\exp\left[-\frac{k^\mu u_\mu}{T_s}\right]\,.
\ea 

\noindent
The two-photon distribution results in 
\ba
J(k,q)_{(\pm)} &=& \int d^4 x\, e^{-i  q \cdot  x} S({x}, k) 
\exp\left[ \mp \frac{\kappa \cdot u_s}{2T_s} \right]
\nonumber\\
&=& 2\,\pi\,R^2\,\gamma_s\, \left(k^\mu \hat\sigma_\mu\right)\,\frac{D}{C_\gamma} \exp\left[-\frac{k \cdot u_s}{T_s}\right] 
\exp\left[ \mp \frac{\kappa \cdot u_s}{2T_s} \right] 
\,\exp\left(-\frac{R^2}{2} q^2\right) \,.
\ea

\noindent
When calculating $R(k,q)$, in the $J(k,q)_{(+)} J(k,-q)_{(-)}$ product 
the terms $\exp[\pm \kappa \cdot u_s /(2T_s)]$ cancel each other.
In the formulae the $k$ and $\kappa$ are
considered as the wavenumber vectors.

We then insert these equations into equation (\ref{Cmain-ss}) and we get for one 
moving Gaussian source
\begin{equation}
C( k,  q) =
1 + \exp\left(-R^2 q^2\right) \ .
\label{Cone}
\end{equation}
Again, this result does not depend on $ k$, just as the 
previous single source at rest in Eq.~(\ref{Csss}).


\begin{figure}[!ht] 
\begin{center}
\includegraphics[scale=0.3]{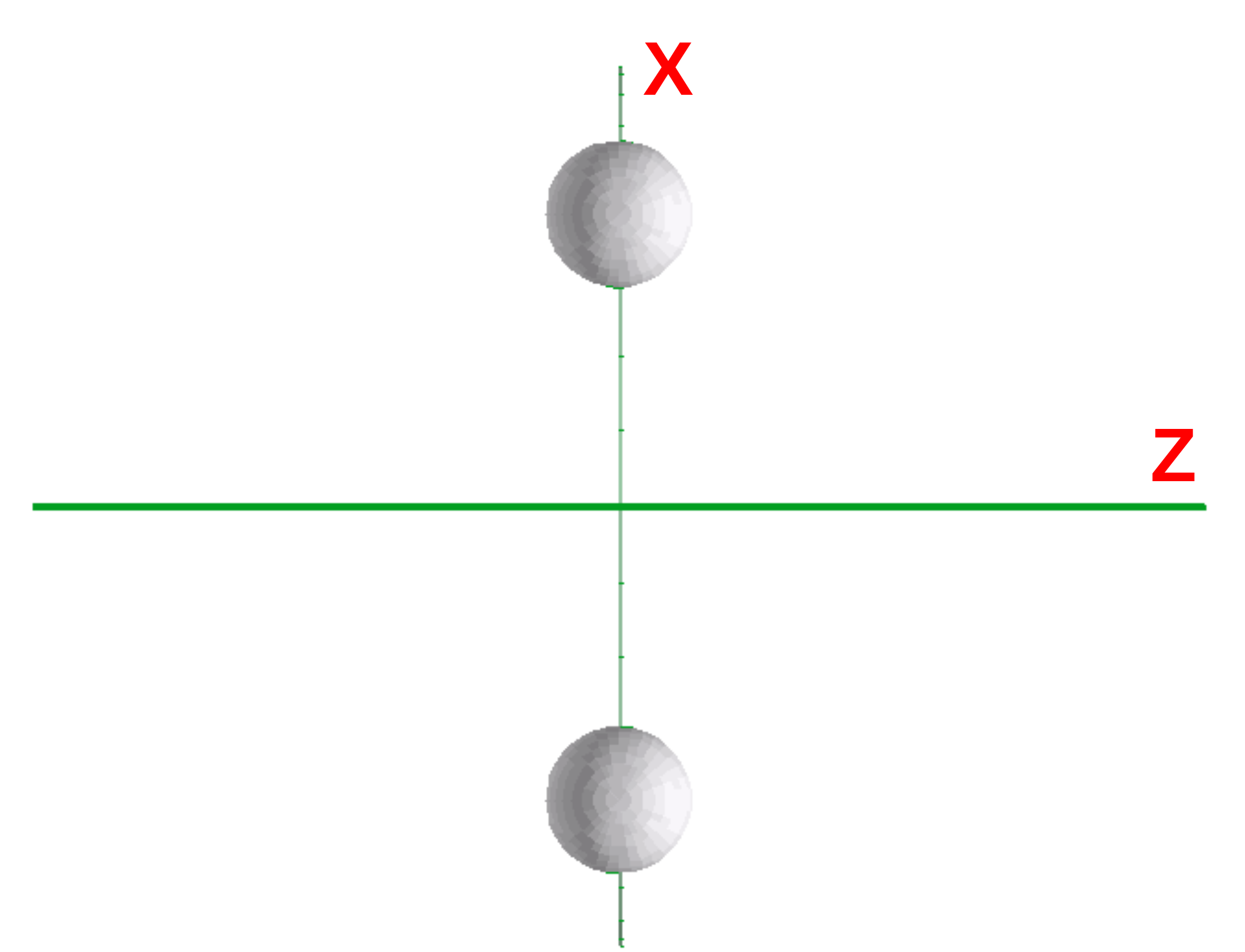}
\end{center}
\vskip -4mm
\caption{
Two steady sources with a distance
between them of $2d$ in the $x-$direction.}
\label{F-2}
\end{figure}

\section{Two Sources}\label{FSM}

\subsection{Two Sources at rest}

For emission from two steady sources in Heavy-Ion Collisions,
two-photon correlations were studied in 
Ref.~\cite{Cso-5}. Here we use our represented method. We assume that 
the two source system is symmetric, i.e. both positions of the sources are placed
symmetrically and 
also their normal vectors, $\hat\sigma^\mu$, are the same, Fig.~\ref{F-2}.
If the normal were $\hat\sigma^\mu =(1,0,0,0)$, then the
invariant scalar $k^\mu\,\hat\sigma_\mu = k^0$; it has been mentioned already    
that we actually do not need this additional requirement to illustrate
the correlation function, which would arise from an idealized symmetric 
system. In case of two symmetric sources at rest ($\gamma_s=1$) the source function is 
(for stars we can savely assume $H_1\left(t\right)=H_2\left(t\right)$): 
\ba 
S(x, k) &=& \sum\limits_{s=1,2}\, S_s\left(x,k\right) 
\nonumber\\
&=& (k^\mu \hat\sigma_\mu) H\left(t\right)\,\sum\limits_{s=1,2} \frac{G_s\left(x,y,z\right) \,}{C_{\gamma}} \, 
\exp\left[-\frac{k \cdot u_s}{T_s}\right] \,,
\ea 

\noindent
while the function $J$ in the J{\"u}ttner approximation is
\ba 
J_{(\pm)}(k,q) &=& \sum\limits_{s=1,2} 
\exp\left[ \mp \frac{\kappa \cdot u_s}{2T_s} \right]\, \exp(iqx_s)
\int_S d^4x\ S_s(x,k)\,\exp(iqx)\,,
\label{J2s} 
\ea 

\noindent
where $x_s$ is the position of the center of the source, 
and the spatial integrals run separately for each of the 
identical sources, i.e. we assume stars with identical density
profiles, but with different temperatures, $T_s$. 

In case of steady sources $u^\mu_s = (1,0,0,0)$, and the spatial integral 
for one source is the same as for a single source. Thus,
\ba 
\int d^4 x\,S\left((x, k\right) &=& \sum\limits_{s=1,2} \int_s d^4 x\,S_s\left(x,k\right) 
\nonumber\\ 
&=& \left(2 \pi R^2 \right) \, 
(k^\mu \hat\sigma_\mu) \sum\limits_{s=1,2}\,\frac{D}{C_{\gamma}} 
\exp\left(-\frac{E_k}{T_s}\right), 
\label{S2sm}
\ea  

\noindent
and 
\ba 
J_{(\pm)}(k,q) &=& 
\sum\limits_{s=1,2} \exp\left[\mp\frac{{\kappa}^0}{2 T_s} \right]\, \exp(iqx_s)
\int_S d^4 x\,S_s\left(x,k\right)\, \exp(iqx)  
\nonumber\\
&=& \left(2 \pi R^2 \right) \, (k^\mu \hat\sigma_\mu) \,  
\exp\left(-\frac{R^2}{2}\,q^2\right) 
\nonumber\\
&& \times \sum\limits_{s=1,2} \frac{D}{C_{\gamma}}
\exp\left(-\frac{E_k}{T_s}\right) 
\exp\left[\mp\frac{{\kappa}^0}{2 T_s} \right]\, 
\exp(i\,q^0\,x^0_s) \exp(-i \vec{q} \cdot \vec{x}_s)\,.
\label{J2sM} 
\ea 

\noindent
In the $J(k,q)_{(+)} J(k,-q)_{(-)}$ product the terms $\exp[\pm {\kappa}^0 /(2T_s)]$
cancel each other. Both $J(k,q)_{(+)}$ and $J(k,-q)_{(-)}$ includes a sum
$[\exp(i\vec q \cdot \vec x_s) + \exp(-i\vec q \cdot \vec x_s)]$, 
and their product leads to a
factor $2[ 1 + \cos(2 \vec q \cdot \vec x_s)]$. 
Consequently, if the two sources have the same parameter, just
different locations, $x_1 = -x_2$ 
(see Fig.\ref{Csss}) then
\be
C(k,q) = 1+ \frac{1}{2} \exp(-R^2 q^2) [1+ \cos(2 \vec q \cdot \vec x_s)]. 
\label{C2}
\ee

\noindent
Like above in case of one source, one may compare the correlation function for binaries in (\ref{C2}) 
with the previous result Eq.~(\ref{C_Binary_10}). The difference between these both results is caused  
by the Gaussian source function in (\ref{Gaussian_Profile}) which is used in order to obtain (\ref{C2}), 
while in (\ref{C_Binary_10}) the Einstein-Hopf model for the thermal radiation source has been applied.  

This result agrees with Ref. \cite{Cso-5} (Section 9.1 on p.41 ibid.),  
and in the limit of $\vec x_s = 0$ it returns the single source
result, Eq.~(\ref{Csss}). 
If the distance of the two sources is 
$2d$, i.e. $x_1=d$ and $x_2=-d$, then $2 \vec q \cdot \vec x_s = 2 q_x\, d$,
thus the modification appears in the $q_x$-direction only. In the
other directions, $q_y$, the single source result
(\ref{Csss}) is returned. For distances under $4R$ the zero-points 
are artificial because the stars will overlap.

\subsubsection{Sources with varying distances}

Eq.~(\ref{C2}) is valid when the two stars are close to each other relative 
to us. That means, they have just past each other in the orthogonal plane. We 
parametrize the distance between the stars, $\vec{x}_s$ with a time component, 
$x_s \rightarrow x_s(t)$. 
%
%
\begin{figure}[h]  
\begin{center}
\resizebox{0.495\columnwidth}{!}
{\includegraphics{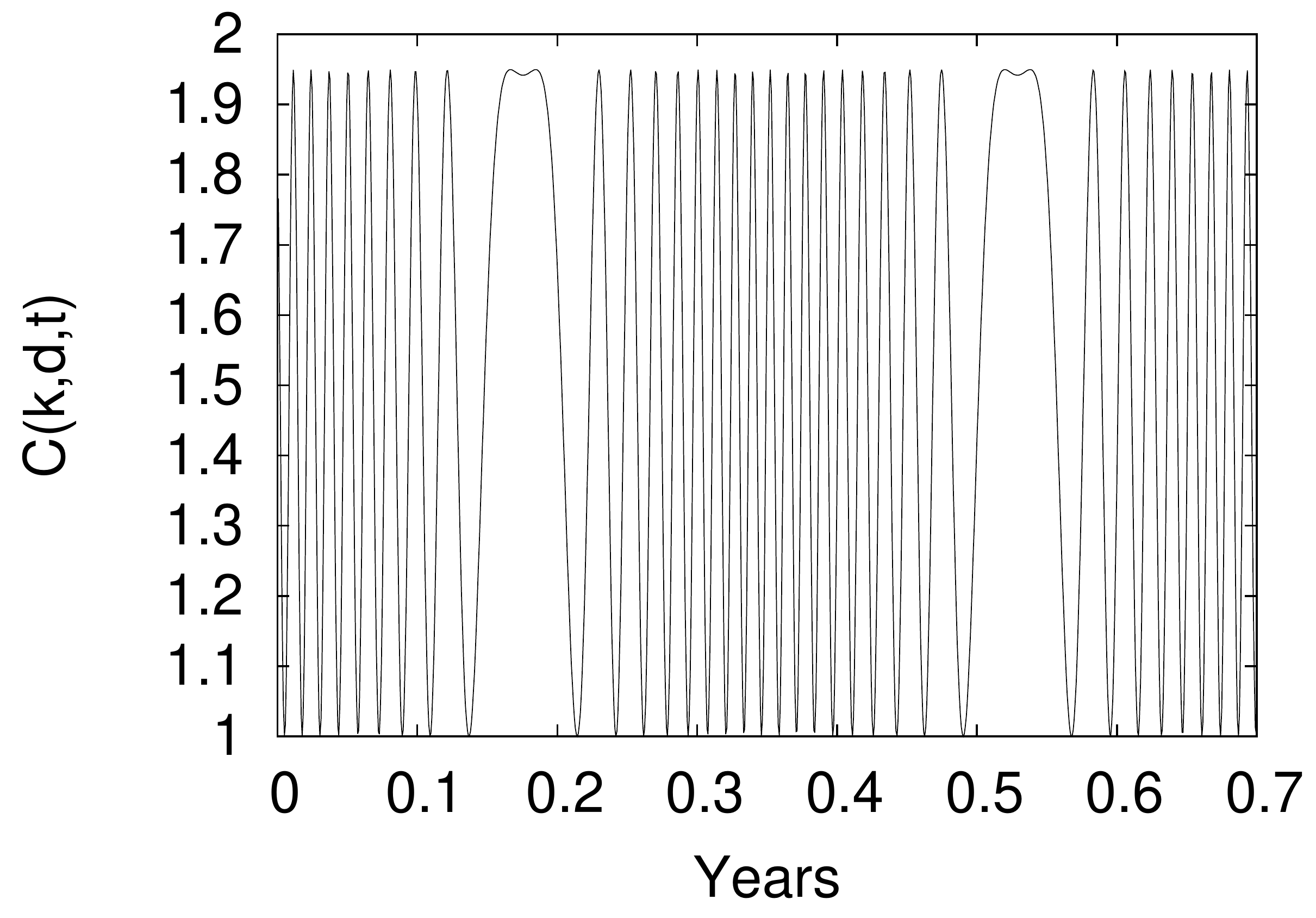}}
\hfill
\resizebox{0.495\columnwidth}{!}
{\includegraphics{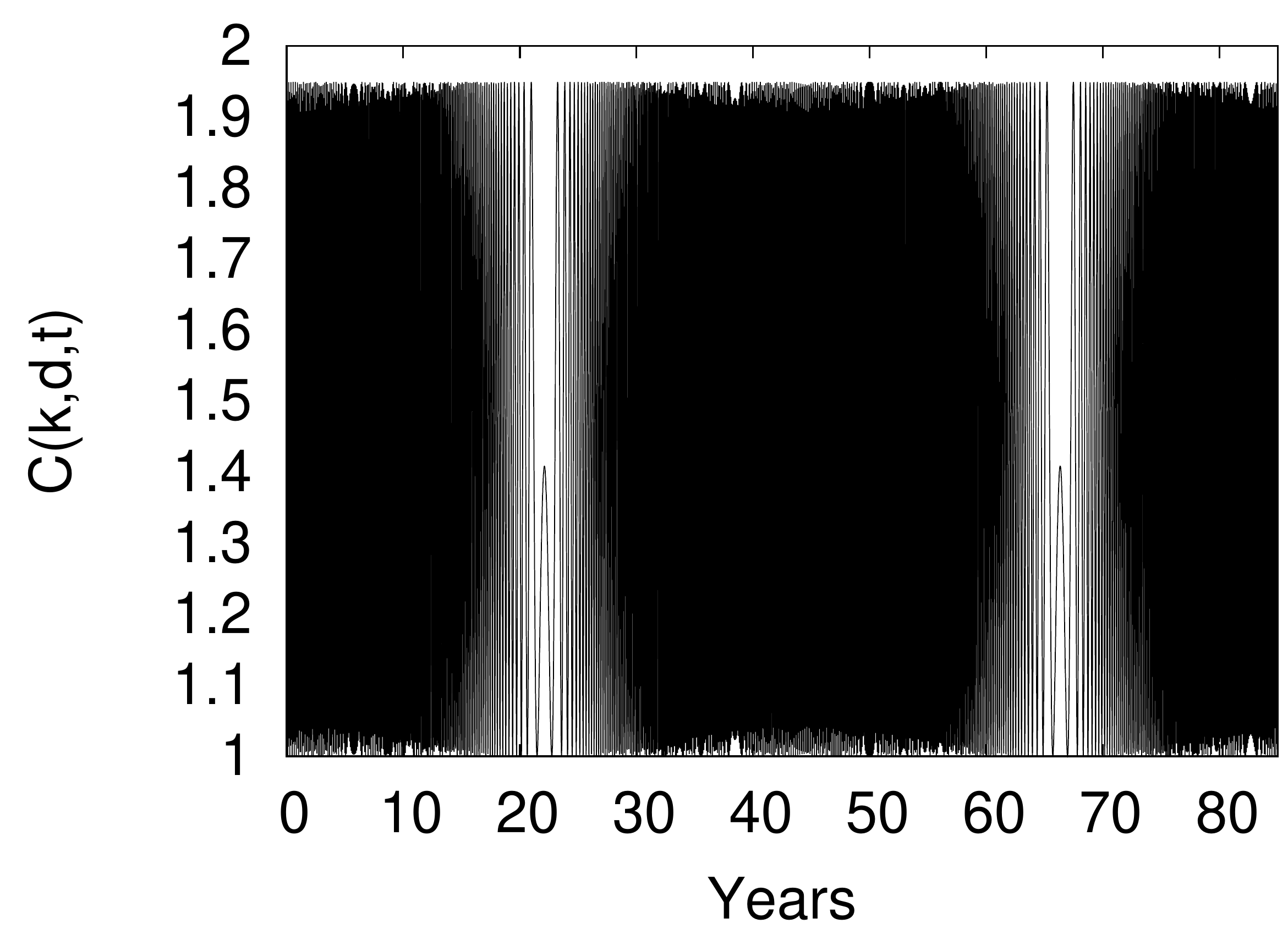}}\\
\resizebox{0.495\columnwidth}{!}
{\includegraphics{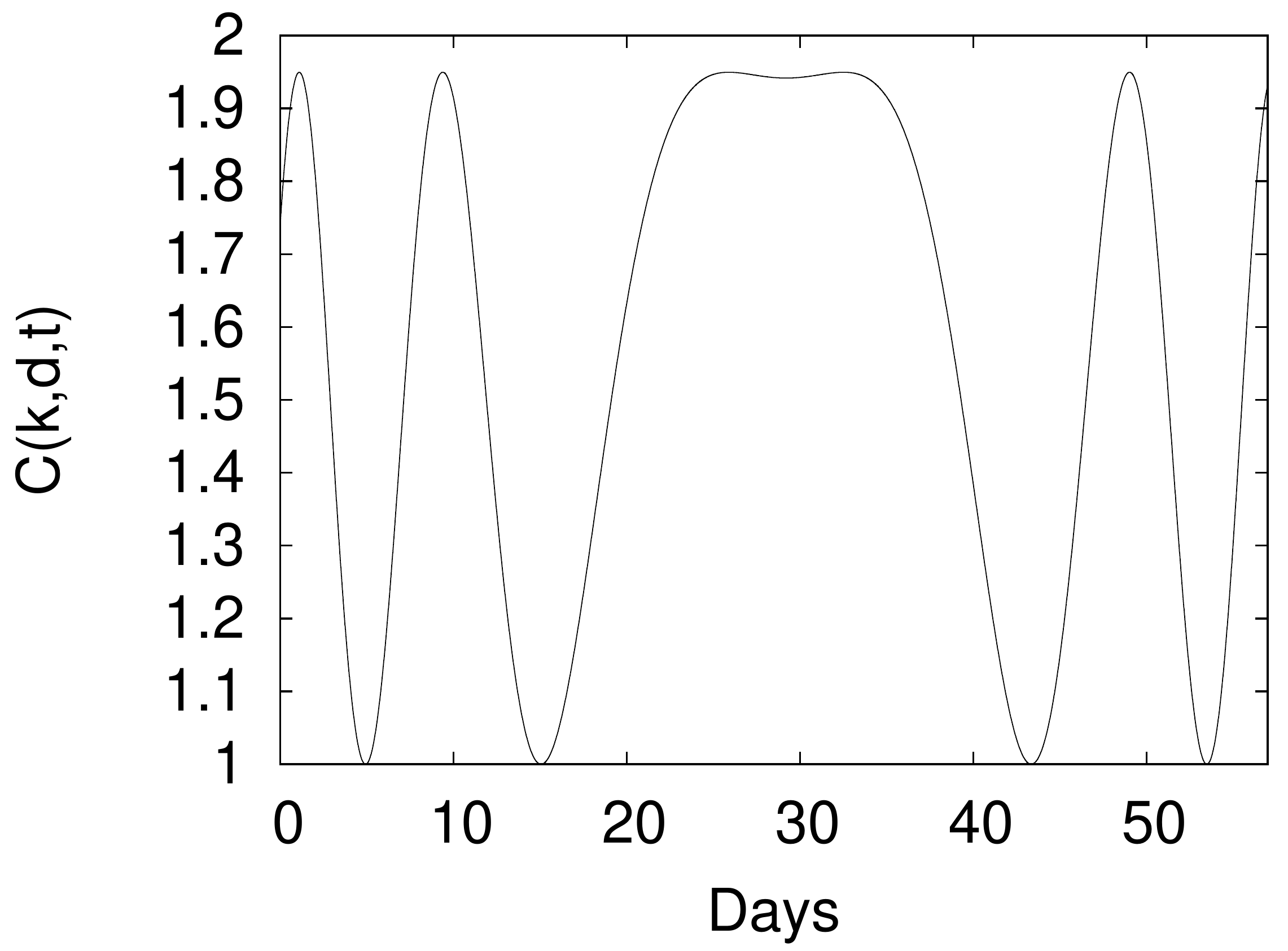}}
\hfill
\resizebox{0.495\columnwidth}{!}
{\includegraphics{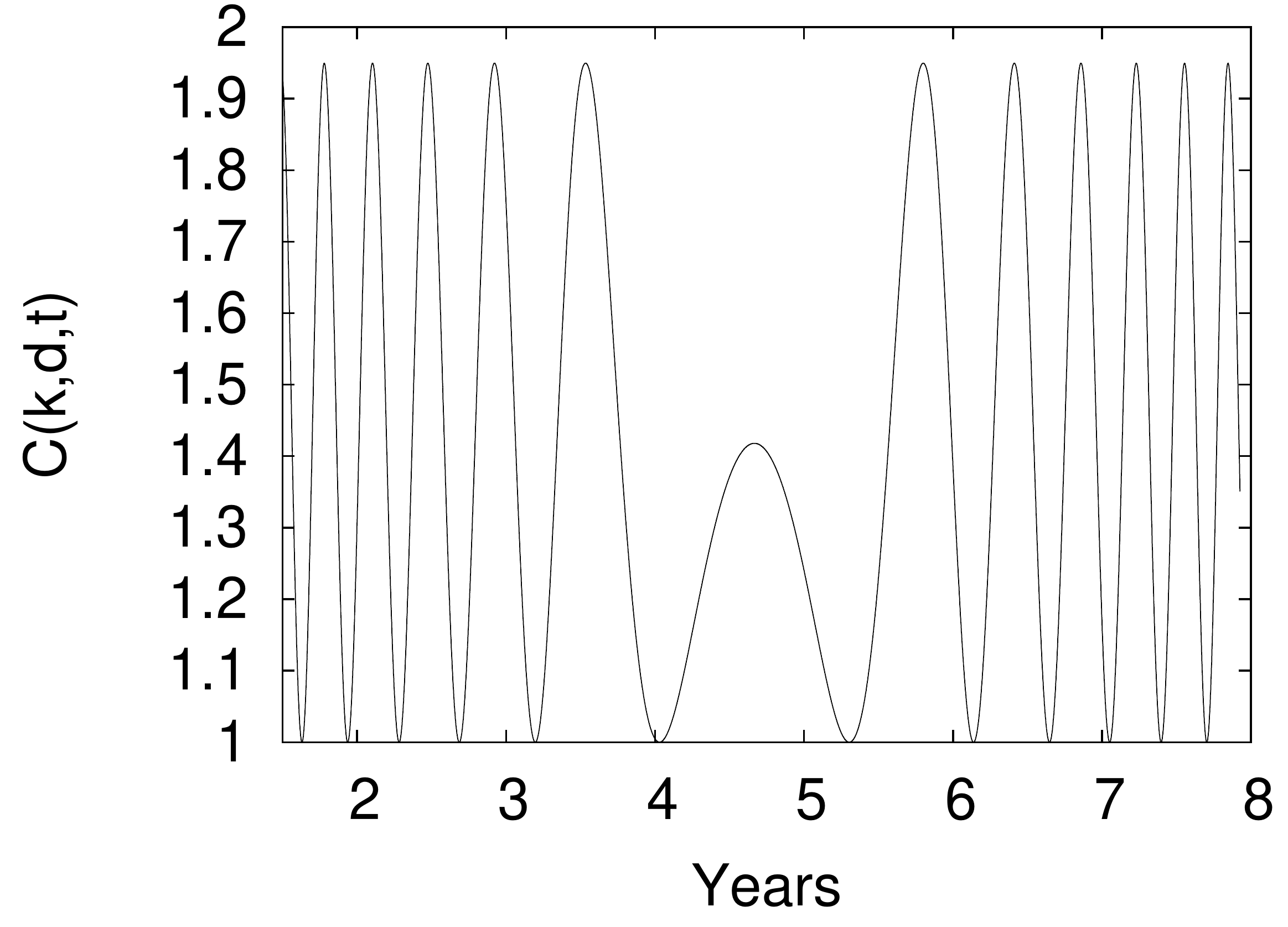}}
\caption{Four diagrams of the correlation function in Eq.~(\ref{C2dt}).  
The two figures at the top show the oscillation of the correlation function over periods 
$1\,{\rm a.u.}$ and $25\,{\rm a.u.}$ respectively.  
The two figures at the bottom show the reduction of the oscillation. The x-axis is to get a perspective of 
the time scales. The parameters are: $\textit{R} = 1.2 \times 10^9$m, $\textit{k} =7.85 \times 10^6$m$^{-1}$, 
$\textit{L}=4.14 \times 10^{16}{\rm m}$, $d=1m$, major axis A is $1\,{\rm a.u.}$, orbital period 0.7 year.} 
\label{F-four}
\end{center}
\end{figure}

Assuming the stars rotate each other in circular 
orbits, the distance $\vec{x}_s\left(t\right)$ on the major axis is given by 
\be
x_s(t) = A\sin(\frac{2\pi}{\mathcal{T}}t+\Phi),
\label{distance1}
\ee
where \textit{A} is the maximum distance of the major axis, and $\mathcal{T}$ 
is the total orbital period. 
$\Phi$ is a phase shift to be adjusted such that when $t=0$, the two stars 
have just past each other, $x_s(0)=2R$, where \textit{R} is the radius of the 
star. Our simple model is not applicable when one star is behind the other, 
shadowing the light from one of the stars. This will give us the possibility 
to have the detector distance fixed, and Eq.~(\ref{C2}) will depend on time.
For two sun-like stars, with $\textit{A=25\,a.u.} $, \textit{G} is Newton's gravitational constant, 
the period is given by Kepler's third law:  
\be
\mathcal{T}=2\pi \sqrt{\frac{A^3}{G(M_A+M_B)}}=2.7 \times 10^9\,s\ \simeq 85\,y, 
\ee
where $M_A$ and $M_B$ are the masses of the two stars.
Hence, Eq.~(\ref{C2}) becomes 
\begin{eqnarray} 
C(q,t) &=& 1+\frac{1}{2}e^{-R^2q^2}[1+\cos(2qx_s(t))] 
\nonumber\\
\nonumber\\
& =& 1+\frac{1}{2}e^{-R^2q^2}[1+\cos(2qA\sin(\frac{2\pi}{\mathcal{T}}t+\Phi))]\,.  
\label{C2dt}
\end{eqnarray}  

\noindent
At $d=1m$ there is a rapid oscillation, but there is a rather special 
behaviour when the sign of the gradient of the sine changes. Four figures 
for two different orbital periods (and hence major axis) are plotted in 
Fig.~\ref{F-four}, one for $A=1\,a.u.$, and the second for $A=25\,a.u.$ respectively.

\begin{figure}[ht]  
\begin{center}
\resizebox{0.45\columnwidth}{!}
{\includegraphics{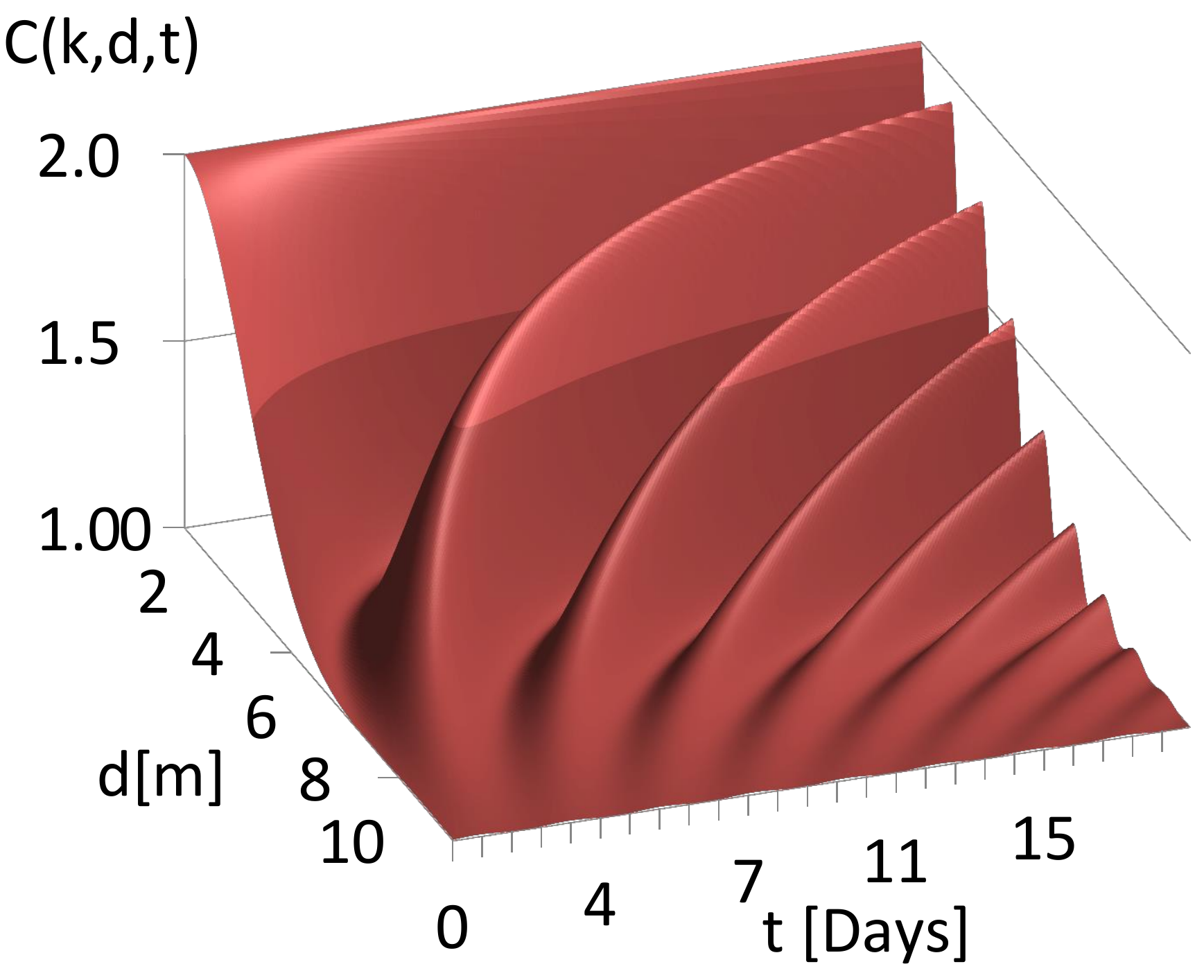}}
\resizebox{0.495\columnwidth}{!}
{\includegraphics{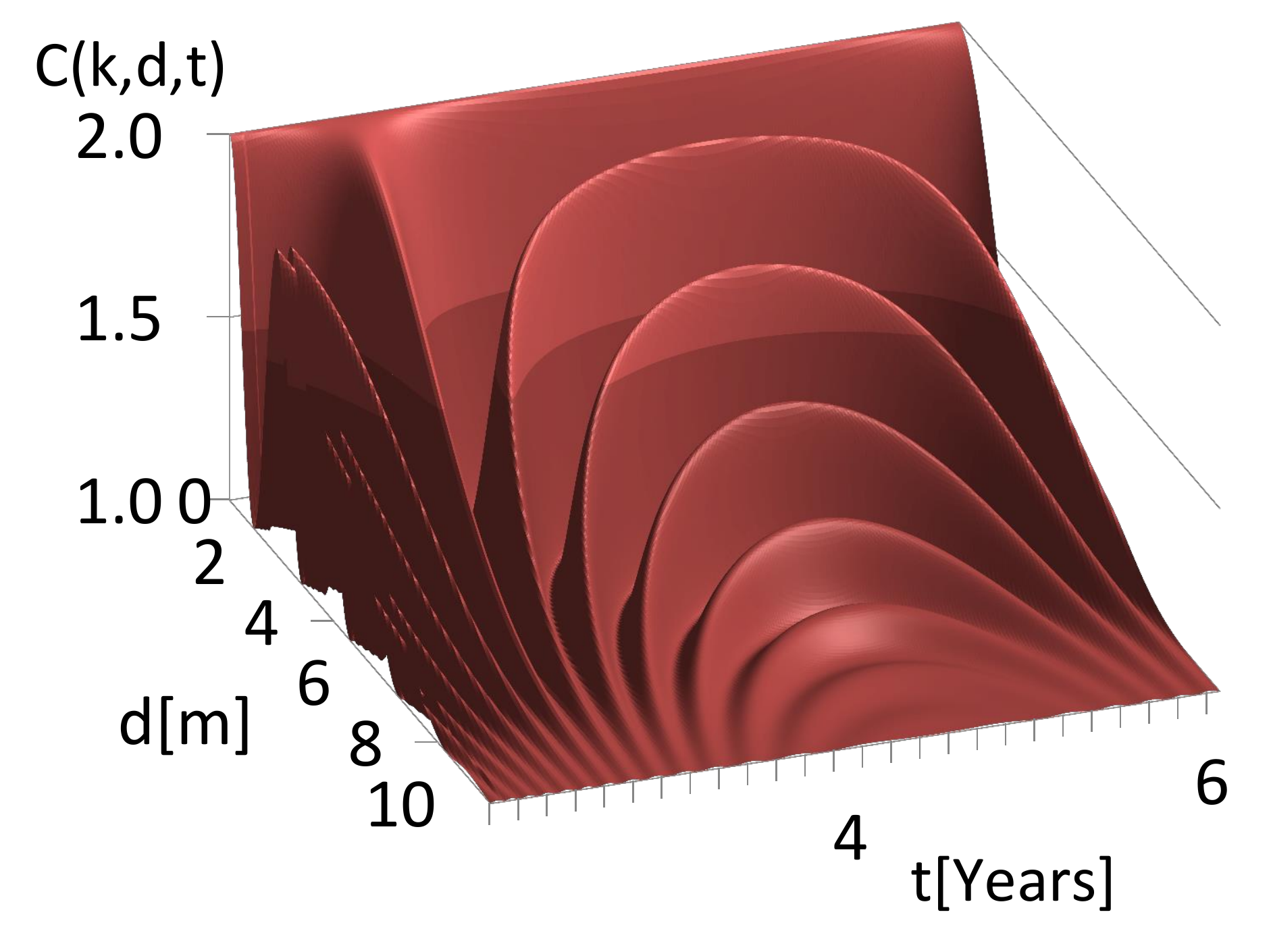}}
\caption{
The correlation Eq.~(\ref{C2}) plotted against time $t$, and detector distance 
$d$, with the corresponding parameters, distance to system 
$L=4.14\cdot10^{16}m$, wave-vector $k=7.85\cdot10^6$, phase shift 
$\Phi = 1.64\cdot10^{-4}$, max major axis amplitude $A=25AU$,  mean-radius of 
stars $R=1.20\cdot10^9m$. Second picture is over a turning point.
}
\label{F-3D}
\end{center}
\end{figure}

As seen in Fig.~\ref{F-four} there are rapid oscillations, but there is 
an interesting effect appearing during a period of rotation.  
During the turning points of the sine function 
the correlation function get a reduction in the oscillation. The reduction appears 
when the stars reaches their maximum separation on the axis parallel to the rotation. If the stars are rotating in 
elliptical orbits, the reduction will be different depending if they are at a 
maximum, or minimum separation. 
Finally, the correlation function over a longer period and over 
different distances is shown in Fig.~\ref{F-3D}. 

\subsubsection{Sources with elliptical orbits}

\begin{figure}[!hb]
\begin{center}
\includegraphics[scale=0.8]{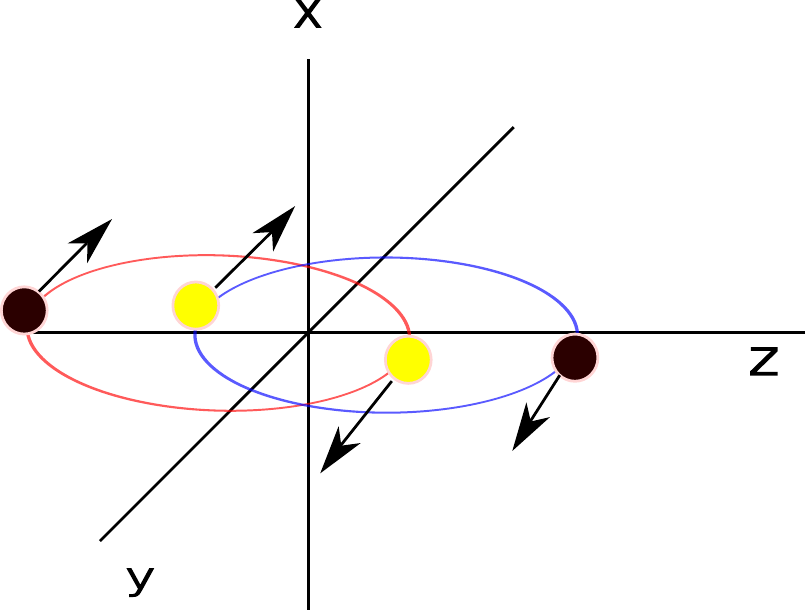}
\end{center}
\caption{ Two Stars orbiting around a common center-of-mass in an 
elliptical orbit.}
\label{elliptical_yz}
\end{figure}

A binary system with an elliptical orbit, Fig.~\ref{elliptical_yz}, is 
characterised by the \textit{eccentricity} $e$. This will effect the relative 
velocities they pass each other as seen from our point of view, if they are 
transiting each other when they are closest, or furthest away. In this ideal 
scenario we will lie on the major axis. Then the stars will move faster when 
they are closest to each other. If we imagine that the two stars pass each 
other when they have their furthest and closest distance, to first 
approximation then they should show two slightly different correlation 
functions after the crossing because of the different velocities. We can 
consider the difference (differential HBT method), 
\be
\delta C(k,q,t) =
\frac{1}{2}
\left(
\cos\left(2q\frac{A}{1{+}e}\sin\left(\frac{2\pi}{\mathcal{T}_1}t+\Phi_1\right)\right)
-
\cos\left(2q\frac{A}{1{-}e}\sin\left(\frac{2\pi}{\mathcal{T}_2}t+\Phi_2\right)\right)
\right)
e^{-R^2q^2},
\label{C-ell}
\ee

\noindent
and this function vanishes for $\epsilon=0$ when they are 
orbiting in perfect circles; here $T_{1/2}$ is the orbital period for the furthest and closest distances,
and $\Phi_{1/2}$ the two phase shits respectively and the dependence on the distance with the
eccentricity is given by $r_{max}=A(1+\epsilon)$ and $r_{min}=A(1-\epsilon)$ \cite{Lien}.
The period are based on
circular motion at these distances, so they are only an approximation for a
short time.

\begin{figure}[ht]  
\begin{center}
\resizebox{0.495\columnwidth}{!}
{\includegraphics{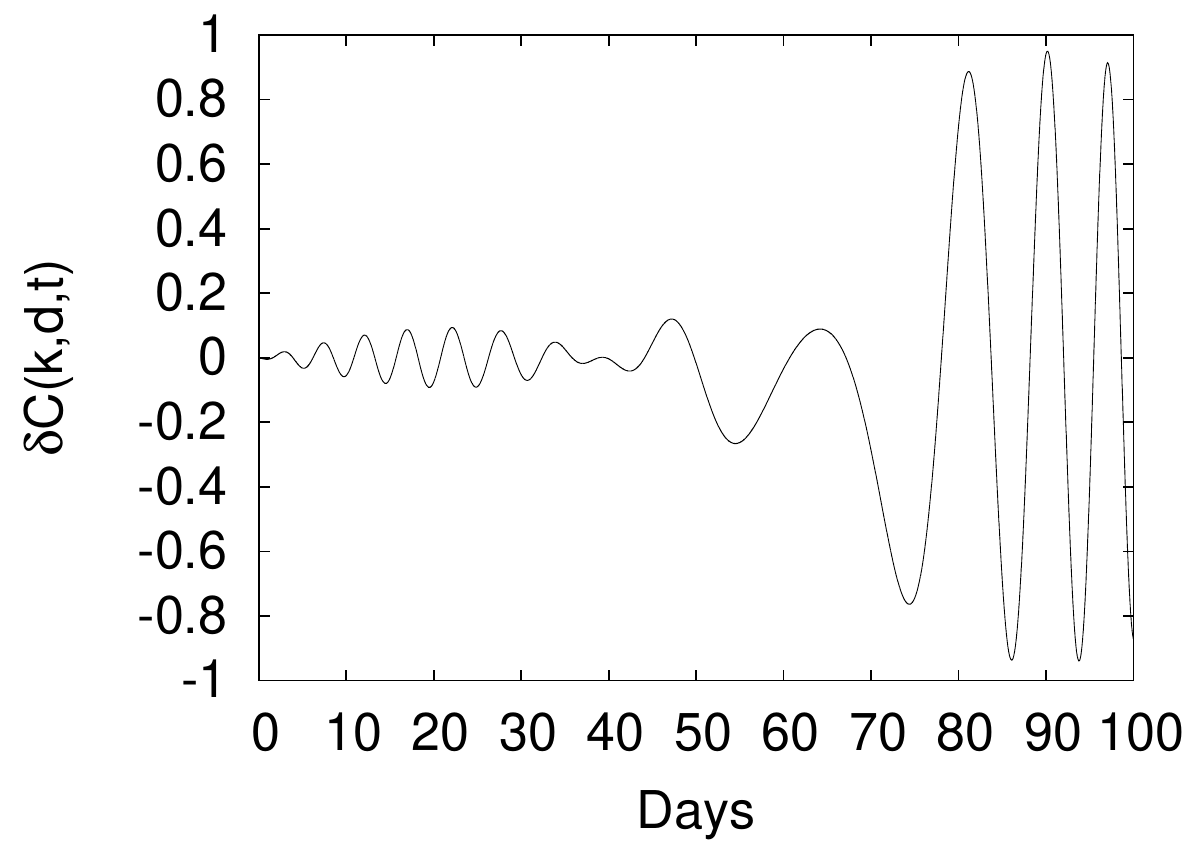}}
\hfill
\resizebox{0.495\columnwidth}{!}
{\includegraphics{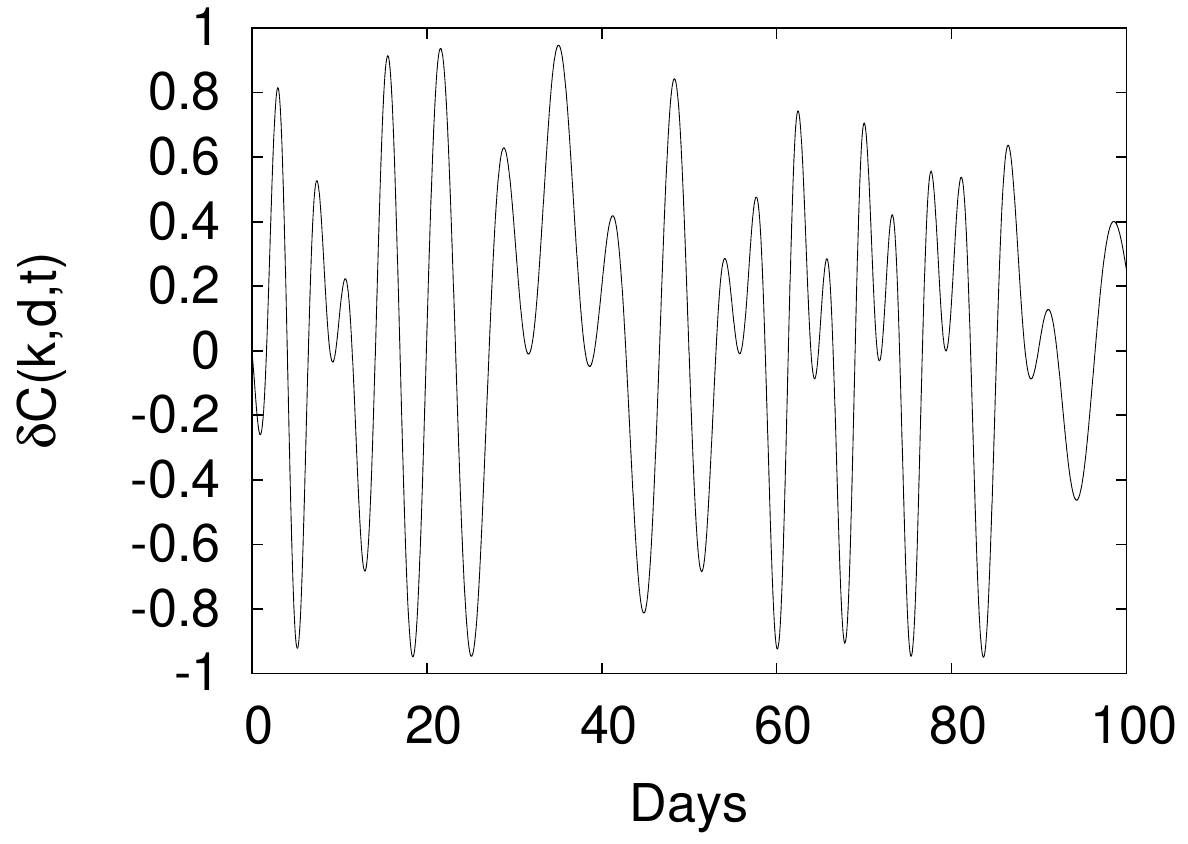}}\\
\caption{The differential HBT correlator as given by Eq.~(\ref{C-ell}).  
The parameter are the following: Left pannel: $e = 0.01$. Right pannel: $e=0.5$.  
Furthermore, orbital period ${\cal T} = 1\,{\rm year}$, 
stellar mean-radius $\textit{R} = 1.2\cdot10^9$m, wave-number $\textit{k} =7.85\cdot10^6$m$^{-1}$, 
distance between stars and detectors $\textit{L}=4.14\cdot10^{16}{\rm m}$, distance of both detectors $d_{AB}=1m$, $A=1\,a.u.$.}  
\label{F-DCF}
\end{center}
\end{figure}

Eq.~(\ref{C-ell}) has a very particular shape that can be seen in Fig.~\ref{F-DCF}a  
and Fig.~\ref{F-DCF}b. It will show its effect in just a 
couple of days. The closer the system is in a circular orbit the longer their 
correlation functions stays in phase. This can be a method of assisting in determination of
the orbital period, velocity and eccentricity in case these two stars can't be resolved 
separately.


\subsection{Two sources in motion}\label{Srs}  

\begin{figure}[!ht]
\begin{center}
\includegraphics[scale=0.3]{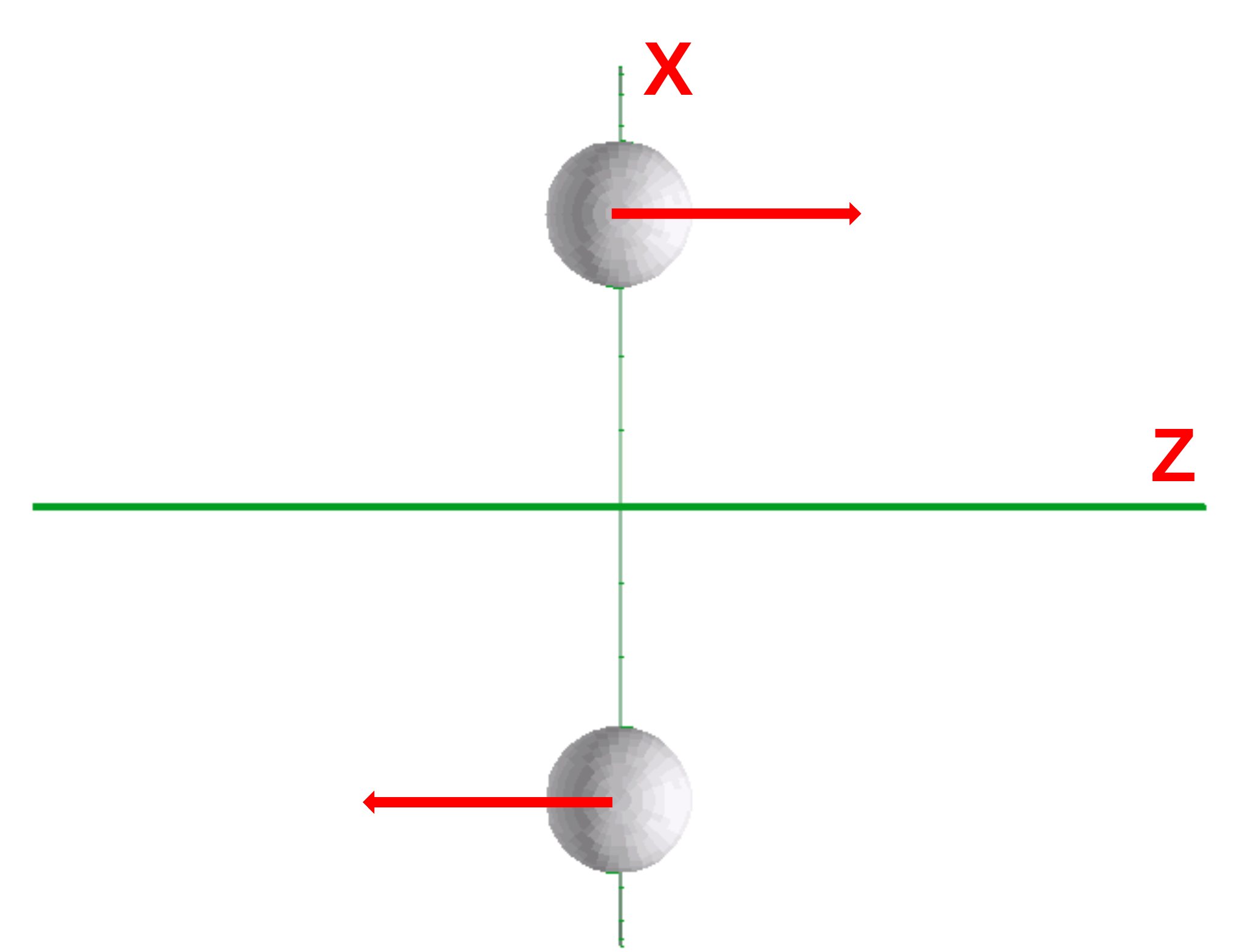}
\end{center}
\caption{ (color online)
Two moving sources in the reaction ($[x-z]$) plane with a distance
between them of $2d$ in the $x-$direction. The sources are moving in the
directions indicated by the (red) arrows.}
\label{F-4}
\end{figure}

We study the system the same way as before, but now we use the present method. The two sources 
are moving in opposite directions, so that
$u_s = u_1$ or $u_2$ where
$u_1^\mu = \gamma_s \left(1, \vec v_1\right)$, 
$u_2 = \bar u_s^\mu = (\gamma_s, \gamma_s (- \vec v_1))$,   and
$\vec u_s \equiv \gamma_s\, \vec v_s$,      so that
$\vec u_1 = - \vec u_2$, see Fig.~\ref{F-4}. Similarly,
$x_s = x_1$ or $x_2$ where
$x_s^\mu = (t_s, \vec x_s)$, 
$\bar x_s^\mu = (t_s, - \vec x_s)$,
and $\vec x_1 = - \vec x_2$.  
We assume again $\hat\sigma^\mu = (1, 0,0,0)$ for the normal of hypersurface, and $t_1 = t_2$.
For binary stars the change of the velocity, and position 
due to the rotation (which is having a frequency of the order of 1/day)
can be considered as quasi-static, so the two-photon correlation
is not effected by this change.  
If we have several sources then the source function is   
\ba 
S\left(x,k\right) &=& \sum\limits_{s=1,2} S_s \left(x,k\right) =  
(k^\mu\,\hat\sigma_\mu) H\left(t\right)\,\sum\limits_{s=1,2} \frac{G_s(x,y,z)}{C_{\gamma}} \, 
\exp\left[-\frac{k \cdot u_s}{T_s}\right] \,,
\ea 
\noindent
while the $J$ function is
\ba 
J(k,q)_{(\pm)} &=& \sum\limits_{s=1,2} \exp\left[\mp \frac{\kappa \cdot u_s}{2T_s}\right]  
\exp(iqx_s) \int_S d^4 x\, S_s(x,k)\, \exp(iqx)\,,
\ea 
\noindent
where $x_s$ is the 4-position of the center of source $s$, 
and the spatial integrals run separately for each of the 
identical sources, i.e. we assume stars with identical density
profiles, but with different  
velocities, $u_s$ and temperatures, $T_s$.

The integral for one source is very similar as for the case of one single source. Thus,
\ba 
&&\int d^4 x \,S\left(x,k\right) = \sum\limits_{s=1,2} \int_S d^4 x\, S_s \left(x,k\right) 
\nonumber\\
&=& (k^\mu\, \hat\sigma_\mu) \, \left(2 \pi R^2 \right) \gamma_s\, 
\frac{D}{C_{\gamma}} \exp\left(-\frac{k^0 \gamma_s}{T_s}\right) 
\left[\exp\left( \frac{\vec k \cdot \vec u_s}{T_s} \right)+ 
\exp\left(-\frac{\vec k \cdot \vec u_s}{T_s} \right) 
\right].
\label{Eq_A}
\ea 

\noindent
This result returns Eq.~(\ref{S2sm}) if $u^\mu_s = (1,0,0,0)$.
The function $J_{(\pm)}(k,q)$ becomes
\ba 
J_{(\pm)}(k,q) &=& \sum\limits_{s=1,2} \exp\left[\mp\frac{\kappa \cdot u_s}{2 T_s}\right]\, 
\exp(iqx_s) \int_S d^4 x\, S_s(x,k)\, \exp(iqx)  
\nonumber\\
= (k^\mu\, \hat\sigma_\mu) 
&&\!\!\!\!\!\!\! \!\!\!\!  
\left(2 \pi R^2 \right)  
\exp\left(-\frac{R^2 q^2}{2} \right) 
\sum\limits_{s=1,2} \frac{\gamma_s\,D}{C_{s}} \exp\left[-\frac{k \cdot u_s}{T_s}\right] 
\exp\left[\mp\frac{\kappa \cdot u_s}{2 T_s} \right]\, \exp(iqx_s)  
\nonumber\\
\nonumber\\
\nonumber\\
= (k^\mu\, \hat\sigma_\mu) 
&&\!\!\!\!\!\!\!\!\!\!\! 
\left(2 \pi R^2 \right) 
\exp\left(-\frac{R^2}{2} q^2\right) \frac{\gamma_s\,D}{C_{s}}
\exp\left[\mp \frac{k^0 \gamma_s}{T_s}\right]
\exp\left[\mp \frac{{\kappa}^0}{2} \frac{\gamma_s}{T_s}\right]
\exp(i q^0 x_s^0) 
\nnb
&& \hspace{-2.7cm} \times \left[\exp\left[\frac{\vec k \cdot \vec u_s}{T_s}\right] 
\exp\left[\pm \frac{\vec{\kappa} \cdot \vec u_s}{2 T_s}  \right]
\exp(-i \vec q \cdot \vec x_s) + 
\exp\left[ -\frac{\vec k \cdot \vec u_s}{  T_s}  \right] 
\exp\left[ \mp\frac{\vec{\kappa} \cdot \vec u_s}{2 T_s}  \right]  
\exp(i \vec q \cdot \vec x_s) \right],
\nonumber\\
\label{J2s3M} 
\ea 

\noindent
where the factor $\exp(i q^0 x_s^0)$ can be dropped if the time 
distribution is simultaneous for the two sources,
because then $x_s^0 = 0$. We have again approximated $\vec \kappa$ as a 
constant, $\kappa = k\Theta$. This allows us treat $\kappa$ as a constant 
of the integrals.
This returns Eq.~(\ref{J2sM}) if $u^\mu_s = (1,0,0,0)$. 
Now we can divide the two-photon correlation with the
square of the single photon distribution 
\begin{eqnarray}  
\frac{Re\,[ J_{(+)}(k,q)\, J_{(-)}(k,-q)]}
{\left| \int d^4 x\,  S(x, k) \right|^2} 
&=& \exp( -R^2 q^2)  
\frac{
\cosh\left( \frac{\displaystyle 2\, \vec k \cdot \vec u_s}{\displaystyle T_s} \right) +
\cosh\left( \frac{\displaystyle \vec \kappa \cdot \vec u_s}{\displaystyle T_s} \right) 
\cos( 2\,\vec q \cdot \vec x_s)}
{\cosh\left( \frac{\displaystyle 2\, \vec k \cdot \vec u_s}{\displaystyle T_s} \right) +1}\,. 
\nonumber\\
\label{C2msd}
\ea 

\noindent
Consequently, if the two sources have the same parameter, just
opposite locations with respect to the center, and opposite velocities,  
then the correlation function is

\ba 
C(k,q) &=& 1 + \exp( -R^2 q^2)
\frac{
\cosh\left( \frac{\displaystyle 2\, \vec k \cdot \vec u_s}{\displaystyle T_s} \right) +
\cosh\left( \frac{\displaystyle \vec \kappa \cdot \vec u_s}{\displaystyle T_s} \right)
\cos( 2\,\vec q \cdot \vec x_s)}
{\cosh\left( \frac{\displaystyle 2\, \vec k \cdot \vec u_s}{\displaystyle T_s} \right) +1}\,. 
\label{C2ms}
\ea 

\noindent
This expression returns Eq.~(\ref{C2}) if $u^\mu_s = (1,0,0,0)$, and $C(k,q) = 2$ if $q = 0$.

\subsection{Two simplest configurations} 

We will now consider two configurations I and II, where in case I the orbital,
$[x,y]$ plane is perpendicular to the direction of the observer, $[z]$, 
on the Earth, and II where the Earth is in the plane of rotation, in direction
$z$.  We assume furthermore that the two stars of the binary have the same
mass, temperature and their orbits are identical circular orbits. This is a 
highly simplified configuration compared to the general configuration in 
Fig.~\ref{Orbital_Parameters}.

\subsubsection{I - Orbital plane is orthogonal to the direction of the Earth}

This configuration is shown in Fig.~\ref{Axes-directions-3}.

\begin{figure}
\begin{center}
\includegraphics[scale=0.4]{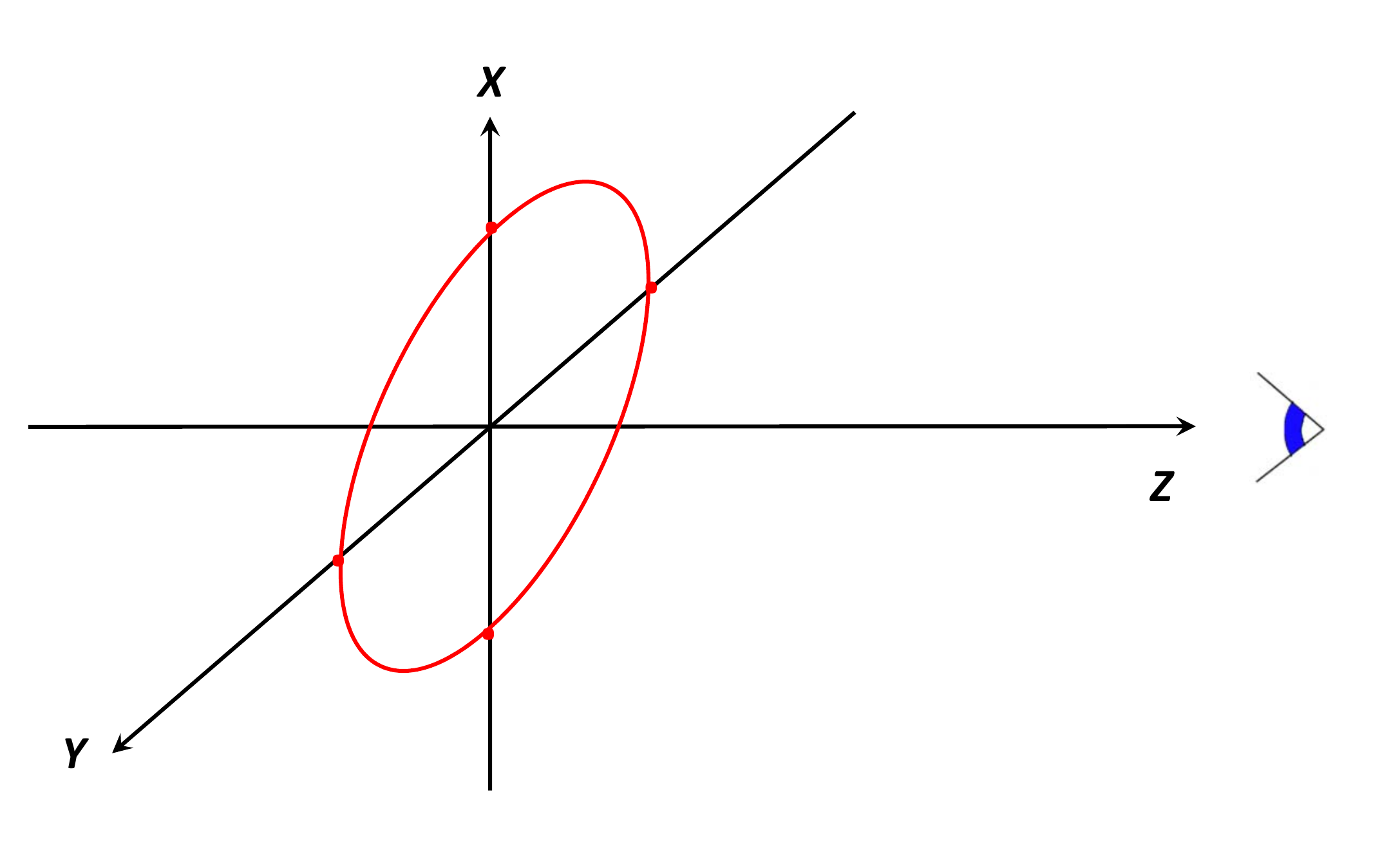}
\end{center}
\caption{The trajectory of the Binary system is in the $[x,y]$-plane,
the observer is in the orthogonal, $z$-direction to the plane. The 
 $z$-axis is the rotation axis.}
\label{Axes-directions-3}
\end{figure}

Let the angular speed of rotation be $\omega$, and $k \gg q$, 
$\vec k = (0,0,k_z)$ but the small differences in $\vec q$ can be measured. 
We need to determine the source parameter, $\vec u_s$ and $\vec x_s$.
If the radius of the orbit is $A$ and the angular velocity of the
rotation is $\omega$ then $u_s = \omega A$ and $x_s =  d = A$. The 
two stars are opposite to each other so that 
$\vec u_A =   \vec u_s$, 
$\vec u_B = - \vec u_A$, 
$\vec x_B = - \vec x_A$.
These vectors are time dependent: So, for anti-clockwise rotation
$\vec x_s = x_s\,( \cos(\omega t), \sin(\omega t), 0)$, and
$\vec u_s = u_s\,(-\sin(\omega t), \cos(\omega t), 0)$.

In case I, the vector $\vec k$ is orthogonal to both $\vec u_A$ and
$\vec u_B$ all the time, so we can have a $z$ directed $\vec k$ only, and so
$(\vec k \cdot \vec u_s)$ vanishes.
The products 
$(\vec q \cdot \vec x_s)$  and  
$(\vec {\kappa} \cdot \vec u_s)$  for $z$ directed $\vec q$ vanish
so
\be
C(k_z,q_z)  = 1 + \exp(-R^2 q^2_z)\ .
\ee

\noindent
Usually $R \ll A$, therefore $C(k_z,q_z)  \approx 2$ for astronomical
configurations, and the dependence on the much larger $A$ leads to stronger
variation at smaller $q$ values:
\ba
C(k_z,q_x)\! &=&\! 1 {+} \frac{1}{2} \exp(-R^2 q^2_x)
\left[1+\cos\left(2 q_x A \cos(\omega t) \right) 
        \cosh\left(\frac{{-} \kappa_x \omega A \sin(\omega t)}{T_s} \right)
\right],
\nonumber
\\
C(k_z,q_y)\! &=&\! 1 {+} \frac{1}{2} \exp(-R^2 q^2_y)
\left[1+\cos\left(2 q_y A \sin(\omega t) \right) 
        \cosh\left(\frac{\kappa_y \omega A \cos(\omega t)}{T_s} \right)
\right].
\label{C2rq}
\ea

\noindent
Here we assumed that the orbital velocity of the binary star components
is non-relativistic, so we can neglect the relativistic $\gamma$ factor. 
$\kappa_x$ and $\kappa_y$ are respectively the mean angular size in x, and y direction. 
For a symmetric source they are both $\kappa_x = \kappa_y = k\frac{2R}{L}=k \Theta$.

As $R\ll A$ the leading exponential term tends to $1$ for small
and intermediate $q_x$ or $q_y$ values. 
At the same time at $q_x = q_y = 0$
both the $\cos$ and $\cosh$ terms tend to $1$, so the value of the 
correlation function is $2$. The $\cos$ terms drop to zero when 
$ q_x  = \pi/[2 A \cos(\omega t)] $ or
$ q_y  = \pi/[2 A \sin(\omega t)] $. These terms
may drop much faster than the leading exponential term as
$A\gg R$, and show a rather special $t$ depencence
due to the orbital rotation. 

\noindent
The arguments in the $\cosh$ terms are of the order $\approx 10^{-8}$ for a binary configuration. 
Same with the two stars orbiting with a distance $1\,a.u.$. This is no surprise because the extra terms 
are coming from the velocity dependence of the emission function through J\"{u}ttner distribution. 
The velocity is given by $\omega A\approx 400\frac{km}{s}$. Thus Eq.~(\ref{C2rq}) effectively goes back 
to the result presented earlier when the velocities of the sources where not explicitly accounted for. 

Eq.~(\ref{C2rq}) goes like the earlier result Eq.~(\ref{C2dt}), as a consequence of the velocity-dependent 
part of the correlation function going to unity. 

\subsubsection{II - The direction of the Earth falls in the orbital plane}

This configuration can in principle occur in two ways, the 
$z$-axis is spanning the distance between the center of the 
Binary and the observer on the Earth, and the plane of rotation 
of the binary can either be the $[x,z]$-plane or the $[y,z]$-plane,
see Fig.~\ref{Axes-directions-2} and Fig.~\ref{Axes-directions-1}.

\begin{figure}
\begin{center}
\includegraphics[scale=0.4]{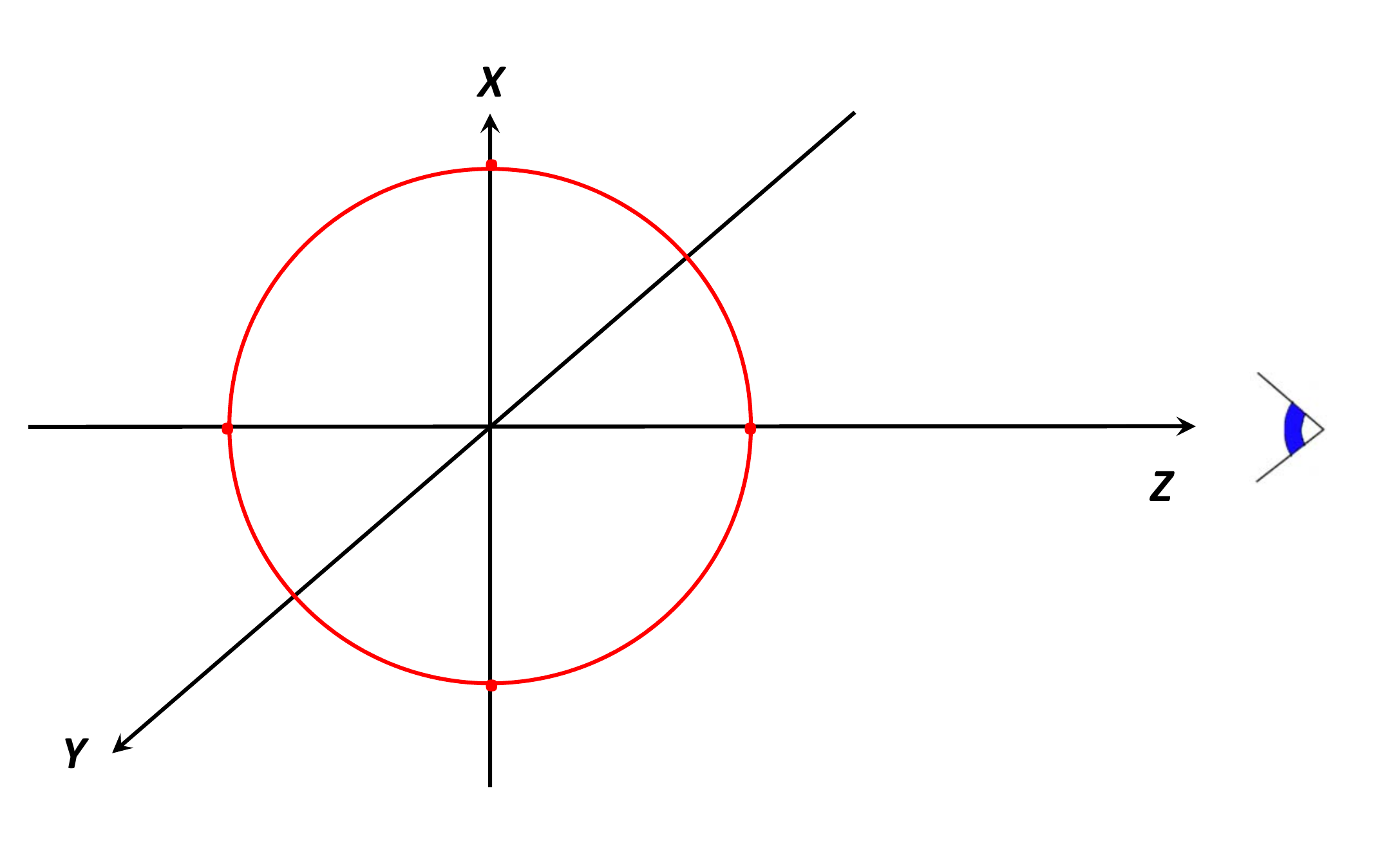}
\end{center}
\caption{The trajectory of the Binary system is in the $[x,z]$-plane,
the observer is in the plane in the $z$-direction.}
\label{Axes-directions-2}
\end{figure}

\begin{figure}
\begin{center}
\includegraphics[scale=0.4]{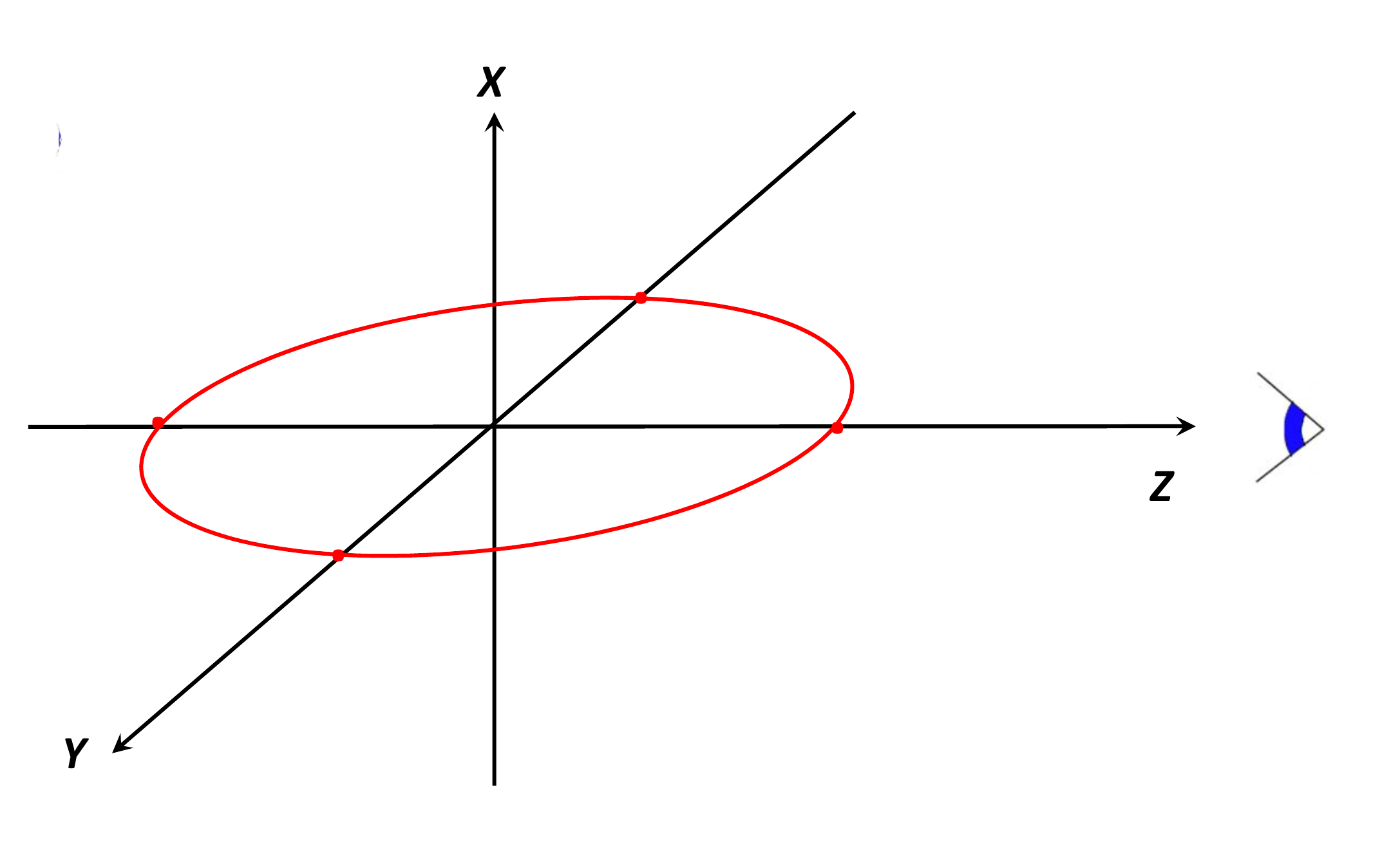}
\end{center}
\caption{The trajectory of the Binary system is in the $[y,z]$-plane,
the observer is in the plane in the $z$-direction.}
\label{Axes-directions-1}
\end{figure}

We can freely choose the direction of the $x$ and $y$ axes, so we chose 
the configuration so that the rotation plane of the Binary is the
$[x,z]$-plane.  Consequently, with this choice neither the distance
between the two starts of the binary nor their velocity will have
any $y$-component, see Fig.~\ref{Axes-directions-2}.
Thus in case II the relation between $\vec{u_s}$ and $\vec{x_s}$ 
are the same, but now the anti-clockwise rotation is in the [x,z] plane,
$\vec{x_s} = x_s\,(\cos(\omega t),0,\sin(\omega t))$ and
$\vec{u_s} = u_s(-\sin\omega t,0,\cos(\omega t))$. 

The decomposition of the correlation function, Eq.~(\ref{C2ms}) now becomes:
\ba
C(k_z,q_x) &=& 1 +\exp(-R^2 q^2_x)
\frac{\cosh\left( \frac{2 k_z \omega A \cos(\omega t)}{T_s}\right)
+ \cosh\left( \frac{- \kappa_x \omega A\sin(\omega t)}{T_s}\right)
\cos(2 q_x A\cos(\omega t))}
{\cosh\left( \frac{2 k_z \omega A \cos(\omega t)}{T_s}\right) + 1}\,, 
\nonumber \\
\nonumber \\
C(k_z,q_y) &=& 1 +\exp(-R^2 q^2_y), 
\nonumber \\
\nonumber \\
C(k_z,q_z) &=&  1 + \exp(-R^2 q^2_z)
\frac{\cosh\left( \frac{2 k_z \omega A \cos(\omega t)}{T_s} \right) + 
\cosh\left(\frac{ \kappa_z \omega A \cos(\omega t)}{T_s} \right) 
 \cos(2 q_z A\sin(\omega t))}
{\cosh\left( \frac{2  k_z \omega A \cos(\omega t)}{T_s}\right) + 1}\,. 
\nonumber\\
\label{configII}
\ea

\noindent
Just as in the previous case,
we have assumed that the orbital velocity of the binary star components
is non-relativistic, so we can neglect the relativistic $\gamma$ factor. 
Also as we have seen the term $\cosh(\kappa_\mu u^\mu) \approx 1$, 
and the argument in $\cosh(k_zu_z)$ is of order $\approx 10^{-4}$.
This is also $1$ for an ordinary binary system. Even for the fastest 
binary observed with a period of 5 minutes the contribution is only of 
$\approx 1.004$ \cite{vik}.

\section{Summary and Outlook}\label{Summary}

Our investigation starts with a reconsideration of the classical HBT effect, both in terms of classical electrodynamics and 
quantum electrodynamics. Within a simple model for the thermal radiation field we have determined the correlation function  
for stars at rest in Eq.~(\ref{Introduction_35b}), and we have explained how one may deduce the diameter of  
stars from this correlator. Afterwards, the correlation function for binary systems has been determined in Eq.~(\ref{C_Binary_10})   
for comparsion with a modified HBT approach used in heavy-ion collisions. Another aspect which has carefully to be treated concerns  
the two-photon wavefunctions, which differ in case of stars and heavy-ion collisions (HIC), because of the different geometrical 
limits of detector distances and radial diamater of the thermal sources, i.e. either stars or hadronic fireballs.  
 
Afterwards, inspired by recent developments with HBT analyses in heavy-ion collisions  
taking the relative velocity of the sources into account, we have tried to  
modify the HBT approach used in heavy-ion collisions to the case of binary star systems.  
Especially, we we have considered the differential HBT approach as it is in use for analyzing the hadronic fireballs created in 
heavy-ion collision experiments. 

Since the fireball in heavy-ion collisions expands rapidly in space and time, 
the correlation function in HIC is determined by covariant integrals over the four-dimensional space-time.
On the other side, stars emit permanently thermal radiation and ther stellar radii remain constant. 
Furthermore, the opacity of stars is usually so extremely high, that the photons only originate from the  
two-dimensional surface of the stars nstead from the entire three-dimensional volume of the stars.   
Accordingly, the HBT approach used in HIC has to be modified for the case of stars by simplifiying 
the four-dimensional integrals itno two-dimensional integrals. 

A further key-point in order to determine the correlator regards the emission function which determines the number of photons 
emitted into a phase-space element. Furthermore, it has been shown that the J{\"u}ttner-distribution, 
frequently used in HIC, is also a good approximation for stars. Then the new approach has been 
applied in order to determine the correlator for stars at rest in Eq.~(\ref{Csss}) and for stars in motion in Eq.~(\ref{Cone}), 
While an exact agreement between the classical approach and the modified HIC approach cannot be expected because 
of the different descriptions for the thermal radiation field,   
we have shown that the result in (\ref{Csss}) corresponds to the previuos result in Eq.~(\ref{Introduction_35b}), 

Then we have determined the correlation function for binary stars at rest to each other in Eq.~(\ref{C2}) and 
for the case of binary stars in motion in Eq.~(\ref{C2ms}). Like in case of one star, we have shown that the result  
in Eq.~(\ref{C2}) obtained from the modified HIC approach corresponds to the expression in Eq.~(\ref{C_Binary_10}) 
which was obtained from the classical HBT approach.  

The first results of our approach have shown that the dependence on 
velocities do not influence the correlation function for normal binary stars, 
as one expected. But the rather specific oscillation dependence of the 
correlation function during a period of rotation is a characteristic sign of 
binary systems, and can help to determine some parameters of the orbits when 
the stars are not optically resolved. This has been elucidated by the difference of the two correlators of a binary system  
(differential HBT approach) in Eq.~(\ref{C-ell}). 

It has been outlined that due to the restriction 
of real experiments being bounded on Earth limits the possibility of fully exploiting the 
correlation function compared to heavy-ion collisions. Our model assumes 
a thermally equilibrium symmetric sources, while more exotic objects could  
in principle also be investigated within the approach presented.  

\section{Acknowledgements} 

One of the authors (S.Z.) thanks for the pleasant hospitality of the
Physics Department at Bergen University (BCPL, Bergen/Norway). He also acknowledges
for kind support from Professor Joerg Aichelin (Subatech, Nantes/France).


\end{document}